\documentclass[%
reprint,
showpacs,
showkeys,
preprintnumbers,
bibnotes,
amsmath,amssymb,aps,prb]{revtex4-1}
\usepackage{graphics}
\usepackage{epsfig}
\usepackage{dcolumn}
\usepackage{color}
\usepackage{bm}
\usepackage{subfigure}
\usepackage{hyperref}
\usepackage[mathlines]{lineno}

\begin{document}


\title{Interplay between Mn-acceptor state and Dirac surface states in Mn-doped Bi$_2$Se$_3$ topological insulator}

\author{M. R. Mahani, A. Pertsova, M. Fhokrul Islam and C. M. Canali}

\affiliation{Department of Physics and Electrical engineering, Linnaeus University, 391 82 Kalmar, Sweden}

\date{\today}

\begin{abstract}
We investigate  the properties of a single substitutional 
Mn impurity and its associated acceptor 
state on the (111) surface of Bi$_2$Se$_3$ topological insulator. 
Combining \textit{ab initio} calculations with microscopic 
tight-binding modeling, we identify 
the effects of inversion-symmetry and time-reversal-symmetry breaking 
on the electronic states in the vicinity of the Dirac point. 
In agreement with experiments, we find evidence that the Mn ion 
is in the ${+2}$-valence state  
and introduces an acceptor in the bulk band gap. 
The Mn-acceptor has predominantly  $p$--character, and is 
localized mainly around the Mn impurity and its nearest-neighbor Se atoms. 
Its electronic structure 
and spin-polarization are determined by the hybridization between 
the Mn $d$--levels 
and the $p$--levels of surrounding Se atoms, 
which is strongly affected by electronic correlations at the Mn site. 
The opening of the gap at the Dirac point depends crucially on 
the quasi-resonant 
coupling and the strong real-space overlap 
between the spin-chiral surface states and the mid-gap spin-polarized Mn-acceptor states.
\end{abstract}

\pacs{73.20.Hb, 
73.20.At, 
71.15.-m, 
}

\maketitle


%

\section{INTRODUCTION}
\label{intro}
Topological insulators (TIs), characterized by a non-trivial insulating gap 
in the bulk and topologically protected helical states on the 
boundaries,  are a new frontier 
in condensed matter physics and materials science~\cite{Hasan,XLQi}. 
Topological surface states (TSSs) in three-dimensional (3D) TIs, 
such as the Bi$_2$Se$_3$ family with a single Dirac cone, 
have attracted particular attention~\cite{zhang2009topological}. 
While the TSSs are robust against time-reversal-invariant 
perturbations, 
the breaking of time-reversal symmetry (TRS) 
opens up an energy gap at the Dirac point. One way to explore 
the response of the TSSs to TRS breaking is via magnetic ordering. 
Apart from being a prerequisite for future spintronic applications, 
the presence of magnetic order in 3D TIs manifests itself in novel quantum phenomena, 
such as the quantum anomalous Hall effect~\cite{Yu02072010,Chang12042013} 
and the topological 
magnetoelectric effect~\cite{PhysRevB.78.195424}.\\
A finite density of magnetic impurities 
on a 3D TI surface  
is expected to bring about a gapped magnetic phase, with magnetic moments 
coupled by a surface-state-mediated exchange 
interaction~\cite{PhysRevLett.102.156603,Checkelsky},    
similar to the carrier-mediated exchange 
coupling in dilute magnetic semiconductors (DMSs). 
However, before one can identify the nature of magnetic interactions, 
it is critical to understand the physics of individual 
magnetic dopants both in bulk and near 
the surface of a 3D TI. 
To date, there seems to be no consensus 
in experimental and theoretical literature 
on the behavior of different species of magnetic impurities in these systems. 
Contrasting results have been reported regarding the chemical trends 
and the magnetic state of the 
impurities~\cite{niu2011mn,henk2012topological,abdalla2013topological,Yuanchang}, as well 
as the presence or absence of the energy gap at the Dirac point upon 
 doping~\cite{Chen06082010,PhysRevLett.108.256810,PhysRevLett.108.117601,honolka2012plane,0953-8984-25-44-445003}.
 For instance, scanning tunneling spectroscopy (STM) and angular resolved photoemission 
   spectroscopy (ARPES) experiments on Bi$_2$Se$_3$ family of 3D TIs 
  have demonstrated the opening of the gap by doping with Fe and Mn~\cite{Chen06082010,Wray2010,PhysRevB.81.195203,Xu2012}. 
   However, more recent work reported striking robustness of the TSS in the  
   presence of magnetic dopants, such as Fe, Co and Gd~\cite{PhysRevLett.108.256810,honolka2012plane,PhysRevB.86.081304,PhysRevLett.108.117601}. 
Theoretical results also differ, with some density functional theory (DFT) calculations confirming the presence of the gap 
at the Dirac point~\cite{li2012strong,zhang2012tailoring,henk2012complex,schmidt2011spin} and others suggesting different scenarios, 
 including a shift at the Dirac cone from the center of the Brillouin zone,~\cite{honolka2012plane, henk2012topological} 
 as well as strong dependence on the magnetization orientation and valence state across the transition-metal (TM) series~\cite{abdalla2013topological}.\\
 We should also note that there exist other possible mechanisms for magnetic ordering in 3D TI systems, different from
 the carrier-mediated (RKKY) interaction~\cite{PhysRevLett.102.156603, PhysRevB.81.233405, PhysRevLett.106.136802, PhysRevB.89.115431} typical of DMSs. 
It is known that in Bi$_2$Se$_3$ both the conduction and valence bands are formed by $p$-orbitals and 
  that the spin-orbit interaction is strong, resulting in band inversion and an energy gap in the bulk. This leads to large matrix elements 
  of the spin-operator between the wave functions of conduction and valence bands and, as a consequence, to a large Van Vleck spin susceptibility~\cite{Yu02072010}. 
  In contrast to DMSs, where this effect is usually small, in the Bi$_2$Se$_3$ family of 3D TIs this can lead to a ferromagnetic order even if the dopants 
   do not introduce free carriers into the host material such as Fe dopants. 
   Hence, possible mechanisms for magnetic interactions may vary depending on the nature of magnetic dopants and 
  different dopant species must be carefully examined.\\
For the important case of magnetic acceptors, 
e.g. Mn on (111) Bi$_2$Se$_3$ surface, a detailed 
microscopic description, 
consistent with experimental observations, 
is lacking. 
There is a strong experimental evidence that Mn behaves 
as a substitutional acceptor 
in the Bi$_2$Se$_3$ family of 3D 
TIs~\cite{PhysRevB.81.195203,APL_152103,phys_20083553}. 
Typically, Mn substitutes Bi in Bi$_2$Se$_3$ in 
the $(d^5)$ configuration, corresponding to the $+2$ valence state, 
giving rise to a spin $S=5/2$.  
Since the nominal valence of Bi is $+3$, 
this implies that substitutional Mn 
impurities also introduce acceptor (hole) states in the  
bulk gap of the host material, similarly to Mn in GaAs, a typical DMS.
These acceptor levels can be directly probed by STM~\cite{PhysRevB.81.195203}.  
However, the nature of these states and their interplay with 
the Dirac surface states 
have not yet been analyzed theoretically.\\
In this work we investigate single substitutional Mn impurities  
on the (111) surface of  Bi$_2$Se$_3$, using 
 DFT  and tight-binding (TB) models. 
We find that 
Mn$^{+2}$ introduces a mid-gap acceptor state, 
localized mainly on the impurity and 
the nearest-neighbor (NN) Se atoms, 
similar to a substitutional Mn in GaAs~\cite{rm_d_level}. 
Our calculations demonstrate the importance 
of electronic 
correlations at the impurity site, which we model by a Hubbard $U$  
parameter~\cite{PhysRevB.44.943}.
The $U$ parameter controls the  position of the 
impurity $d$--orbitals, which in turn determines the hybridization 
with the $p$--orbitals of NN Se atoms and the acceptor spin-polarization. 
Increasing $U$ 
localizes the Mn $d$--states, leading to an enhancement 
in the Mn magnetic moment and a weakening of the $p$--$d$ hybridization
and the acceptor polarization. 
With the Mn placed on one of the surfaces of a finite slab, 
the spin-polarized acceptor states 
couple quasi-resonantly with the helical TSS at the same surface, 
opening a gap of a few meV at the Dirac point. The magnitude of 
 the gap is significantly affected by 
the strength of the $p$--$d$ hybridization. 
 With the appearance of the energy gap,
the system exhibits a finite out-of-plane 
magnetization~\cite{PhysRevLett.102.156603,Checkelsky}.\\
The rest of the paper is organized as follows. 
In Section~\ref{models} we discuss the details of the DFT  
 calculations and the TB model for magnetic and nonmagnetic impurities in Bi$_2$Se$_3$. 
 The results of DFT and TB calculations are presented in 
Section~\ref{results}. In particular, we describe 
modifications in the electronic bandstructure of a  
 Bi$_2$Se$_3$ slab, induced by doping, 
 and analyze the electronic and spin properties 
 of the acceptor states, associated with Mn impurities.  
 The role of the spin-polarized acceptor states in the opening of the gap at the 
  Dirac point is discussed. Finally, we draw some conclusions.
\section{Computational Models}
\label{models}
The DFT calculations were performed using the 
 full-potential all-electron linearized augmented 
plane waves method as implemented 
in the WIEN2k package~\cite{Wien2k_package}.
The generalized gradient approximation (GGA) is used for
 exchange correlation functional~\cite{perdew96}.
We consider a  $2\times2$ surface supercell containing six quintuple layers (QLs) 
of Bi$_2$Se$_3$.  A Bi atom in the topmost Bi monolayer (ML) 
is replaced by a Mn (Mn doping of $~$2\%). 
The direction of the magnetization 
is along [001] ($z$-axis), which is perpendicular to the (111) surface. 
A vacuum of 30 Bohr is added along the [001] direction to avoid supercell interaction.
The atomic positions in 
the supercell have been fully relaxed. 
We use four non-equivalent 
$k$-points in the Brillouin zone. 
Electronic correlations at the impurity site are accounted for by means 
of the (GGA+$U$)-method. 
In Section~\ref{results} we will consider explicitly the two cases $U=0$ and $U=4$~eV~\cite{rm_d_level}.\\
In addition to DFT calculations, to model the electronic structure of pristine Bi$_2$Se$_3$, 
we use the \textit{sp}$^3$ Slater-Koster TB Hamiltonian with parameters 
 fitted to DFT calculations, which has been discussed extensively in 
 Ref.~\onlinecite{kobayashi2011electron} and \onlinecite{ap_TI_prob}. 
 An impurity is introduced in the TB Hamiltonian via a local modification of the on-site potential at the impurity site. 
For a non-magnetic impurity, 
the on-site energy is modified as 
$\tilde{\varepsilon}_{i\alpha\sigma}=\varepsilon_{i\alpha}+\varepsilon^{p}_{i\alpha}$,
where $i$ is the index of the atom where impurity is located,
$\alpha$ is the orbital index and $\sigma$ is the spin; 
$\varepsilon_{i\alpha}$ is the spin-independent on-site energy
of atom $i$ in the pristine case and $\varepsilon^{p}_{i\alpha}$ is a spin-independent potential shift. 
For numerical calculations we choose a value $\varepsilon^{p}_{i\alpha}$=$0.75\,\varepsilon_{i\alpha}$, which  
 generates a shift between the TSSs, corresponding to top and bottom surfaces, of the same order of magnitude as that obtained in our DFT 
calculations.  
In the case of a magnetic impurity, we assume that, apart from an overall potential shift, the impurity 
 induces a local spin-splitting. Therefore the modified on-site energy of the impurity atom is written as $\tilde{\varepsilon}_{i\alpha\sigma}=
\varepsilon_{i\alpha}+\varepsilon^{p}_{i\alpha}\pm\varepsilon^{s}_{i\alpha}$ ($+$ for $\sigma=\uparrow$ and $-$ for $\sigma=\downarrow$; 
the spin-quantization axis is along the $z$-axis, perpendicular to the (111) surface of a Bi$_2$Se$_3$ slab). The spin-dependent part of the on-site potential, 
$\pm\varepsilon^{s}_{i\alpha}$, which is taken to be a fraction of $\varepsilon_{i\alpha}$, 
 enforces a net out-of-plane magnetization and generates a small gap at the Dirac point of the TSS of the doped surface (top in our calculations).  
This crude impurity model allows us to illustrate 
the main features of the electronic bandstructure, associated with inversion symmetry (IS) and TRS breaking.\\
\section{Results and discussions}
\label{results}
\begin{figure}[htp]
\centering
\includegraphics[scale=0.08]{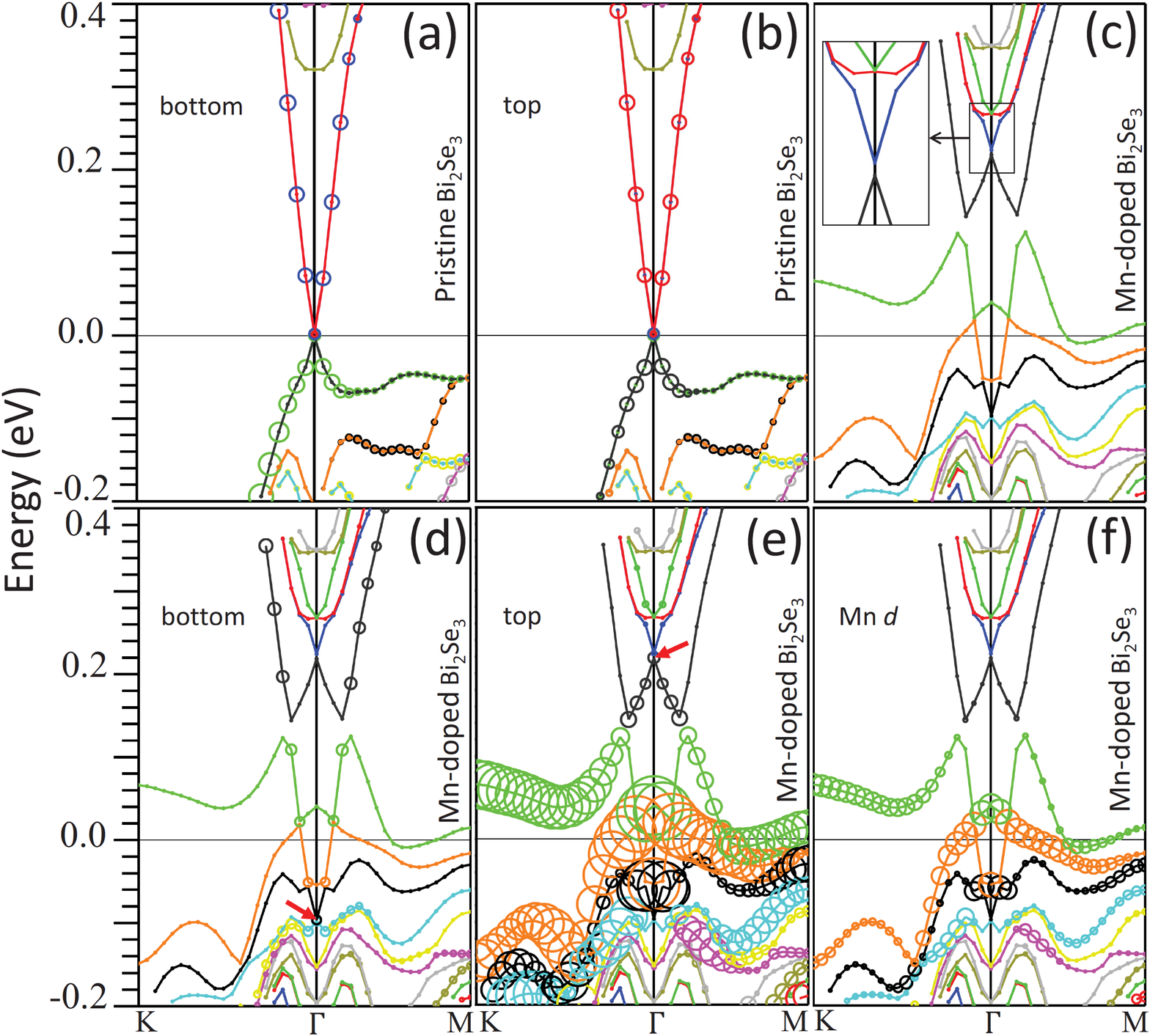}
\caption{Bandstructure of pristine (a,b) 
 and Mn-doped (c-f) Bi$_2$Se$_3$, calculated with DFT for $U$=$0$.  
Inset shows the gap at the Dirac point 
 caused by the TRS breaking. 
  Circles show the contribution 
  of the bottom (a,d) and top (b,e) surface states, 
  and of the Mn $d$--orbitals (f). 
 The radius of the circles is 
proportional to the relative weight  
at a given energy and $k$--point. Arrows in panels 
 (d) and (e) mark the Dirac points of the bottom and top TSS, 
 respectively.}
\label{fig:GGA}
\end{figure}

We start with the DFT bandstructure of a pristine Bi$_2$Se$_3$ slab, plotted 
in Figs.~\ref{fig:GGA}(a)-(b), showing the expected conical TSS crossing at the Dirac point. 
The TSS consists of two degenerate states, 
one for each slab surface. 
 For the particular slab considered, these states are 
only slightly coupled, introducing a small ($\lesssim 1$~meV) gap at the Dirac point. \\
\begin{figure*}[htp]
\centering
\includegraphics[scale=0.5]{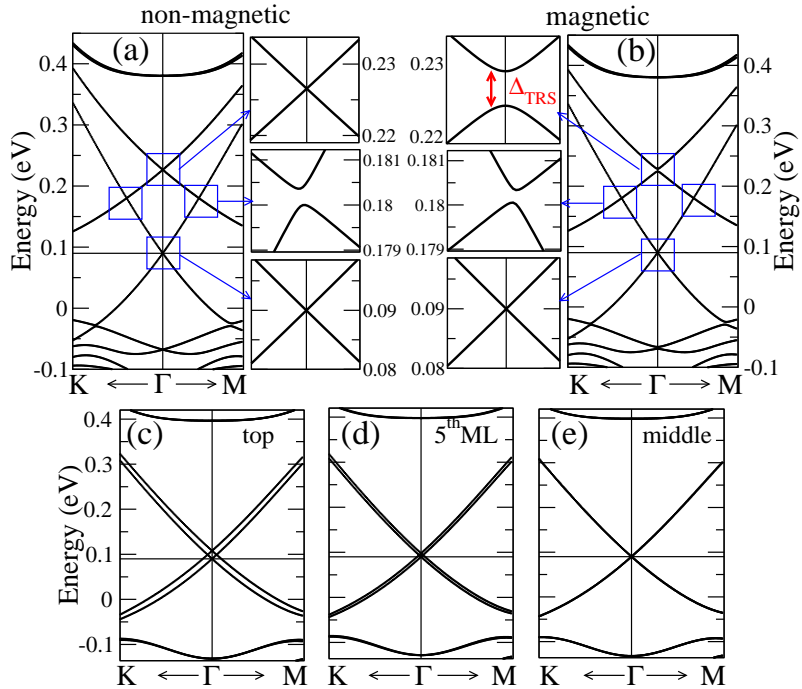}
\caption{(Color online) Bandstructure of Bi$_2$Se$_3$ calculated 
with the TB model using a 3$\times$3$\times$6QL supercell with  
(a) a non-magnetic and (b) a magnetic impurity, substituting a Bi 
atom in the topmost Bi ML. Bandstructure for a non-magnetic impurity 
at different depth in a 3$\times$3$\times$5QL slab: (c) topmost ML, 
(d) $5^\mathrm{th}$ ML below the surface and (e) $13^\mathrm{th}$ ML, corresponding to the middle
 of the slab.  
}
\label{fig:TB_non_mag}
\end{figure*}
The DFT bandstructure of Mn dopants substituted on one (the TOP) of the slab surfaces 
is plotted in Fig.~\ref{fig:GGA}(c), for the case $U=0$. 
It is characterized by the following features: 
(\textit{i}) the conical TSS belonging to the top surface (where the impurity resides) has been pushed up in energy. 
The bottom TSS 
is essentially unaffected by the impurity 
[see also Figs.~\ref{fig:GGA}(d)-(e), where the bottom and top states are highlighted]; 
(\textit{ii}) the two displaced conical TSSs exhibit avoided level crossings, with a gap of the order of $20$~meV,
at two symmetric $k$-points with respect to $\Gamma$;
(\textit{iii}) the Dirac point of the top (bottom) TSS is now
above (below) the Fermi energy ($E=0$). For the top TSS there is an energy gap 
of $\approx 5.5$~meV at the Dirac point [see the inset in Fig.~\ref{fig:GGA}(c)]. As we explain below, this 
gap is caused by the TRS breaking due to magnetic doping. 
For the bottom TSS the gap remains negligible, i.e. $\sim 1$~meV, as in the pristine case;  
(\textit{iv}) the states in the energy window $E\in [0, 0.13]\ {\rm eV}$, 
not belonging to the bottom TSS,  
result from the complex hybridization of Mn $d$--levels, 
NN  Se $p$--levels 
and the extended top TSS.\\
The $\approx 5.5$~meV gap at the Dirac point obtained in 
our calculations matches reasonably well  the $\approx 7$~meV 
gap found experimentally for 1\% Mn doping in Bi$_2$Se$_3$~\cite{Chen06082010}. 
The energy gap is expected to increase for higher doping concentrations.\\
In order to explain some of these features, 
we employ the ${sp}^3$ TB model for Bi$_2$Se$_3$, described in Section~\ref{models}.
%
Figures~\ref{fig:TB_non_mag}(a)-(b) show the bandstructure of 6QLs 
 of Bi$_2$Se$_3$ calculated with the TB model, where a non-magnetic or a magnetic impurity 
 is substituting Bi in the second ML below the surface.
In the non-magnetic case, as a result of asymmetric doping, 
one of the two degenerate (for pristine Bi$_2$Se$_3$) conical TSSs, 
 corresponding to the doped surface,  
 is shifted up in energy. 
 The TSS of the un-doped surface 
remain nearly unaffected, as    
 expected for a relatively thick slab. 
 Since the TRS is preserved, the states at $\Gamma$ 
 have a two-fold degeneracy related to the opposite spins [top inset 
  in Fig.~\ref{fig:TB_non_mag}(a)]. However, the asymmetric doping 
   breaks the inversion symmetry (IS), therefore away from $\Gamma$ the degeneracy 
  is lifted and avoided crossings are formed at two symmetric $k$-points, 
 producing a gap, which can be seen in the middle insets in Figs.~\ref{fig:TB_non_mag}(a) and (b)  
  (this gap should vanish in the limit of an infinitely thick slab).  
 The presence of the magnetization 
breaks the TRS, which leads to  the lifting of the 
degeneracy at all $k$--points. Indeed, in the magnetic case, in addition to the features related 
to asymmetric doping, we find a gap at the Dirac point ($\Gamma$) 
 for the TSS of the doped surface [top inset in Fig.~\ref{fig:TB_non_mag}(b)].\\
We focus specifically on three important regions in the 
bandstructure, namely the Dirac point of the top surface 
states [top inset in Fig.~\ref{fig:TB_non_mag}(b)], which interact with 
impurity, the Dirac point of the bottom surface states 
(unperturbed) [bottom inset in Fig.~\ref{fig:TB_non_mag}(b)], and 
 the avoided level crossings, which occur symmetrically on both 
sides of the $\Gamma$ point and introduce a small gap 
[middle inset in Fig.~\ref{fig:TB_non_mag}(b)]. The latter 
feature, as well as the overall shape of the bandstructure consisting of two 
  shifted Dirac cones, is present in both non-magnetic and 
	magnetic cases. As we explain in detail below, it is caused 
	purely by IS breaking since the impurity is positioned on 
	only one of the surfaces of the slab. 
 However, the crucial difference between non-magnetic and 
magnetic impurities is the opening of the gap at 
the Dirac point of the TSS of the top surface, where the impurity is located. 
  This is a manifestation of the TRS breaking. 
	Note that the bottom TSS remain essentially unperturbed by the impurity in both cases.\\
We further investigate the effect of asymmetric doping by placing a non-magnetic impurity 
at different positions in the slab [see Fig.~\ref{fig:TB_non_mag}(c)-(e)]. Note that in this case 
the impurity substitutes Se and we use a 5QL slab; such configuration allows us to position 
the impurity at the exact geometrical center of the slab, which is a Se monolayer (ML). We use a similar potential shift, 
$\varepsilon^{p}_{i\alpha}$=$0.75\,\varepsilon_{i\alpha\sigma}$, as in the 
case of impurity substituting Bi. Placing the impurity close to one of the surfaces 
shifts the corresponding Dirac cone 
while the other one remains unchanged, provided that the slab is thick enough 
so that the two surfaces do not interact strongly with each other [Fig.~\ref{fig:TB_non_mag}(c)].
As the impurity is moved further away from the surface, its interaction with the topological 
 surface states on the corresponding 
surface decreases, reducing the shift of the Dirac-cone states [Fig.~\ref{fig:TB_non_mag}(d)].
Finally, when the impurity is placed in the middle of the slab, the IS is restored and 
we find two degenerate Dirac cones with the position of the Dirac point coinciding with that 
of the pristine slab. 
Based on these calculations, we attribute similar features occurring 
 in the vicinity of the Fermi energy in the DFT bandstructure [See Fig.~\ref{fig:GGA}(c)]  
to the IS breaking caused by asymmetric doping.\\ 
\begin{figure}[htp]
\centering
\includegraphics[scale=0.43]{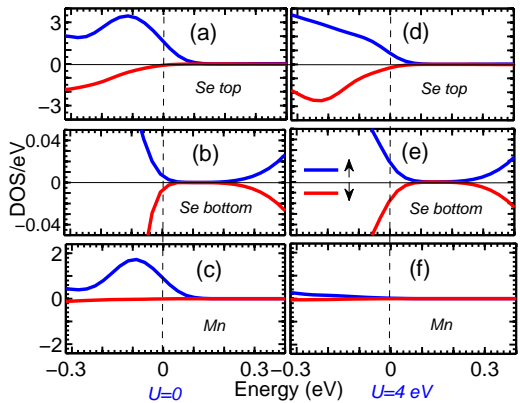}
\caption{Partial $p$--DOS for Se atoms 
in the top two (a,d) and bottom two  (b,e) Se MLs;  
(c,f) $d$--DOS for the Mn impurity. Left panels are for  
$U$=$0$, right panels for $U$=$4$~eV. Positive (negative) 
DOS corresponds to majority (minority) spin. 
The vertical line marks the position of the Fermi level ($E=0$).}
\label{fig:DFT_DOS}
\end{figure}
We now focus on the properties of the unoccupied electronic states, 
appearing above the Fermi level in our DFT bandstructure 
calculations [Figs.~\ref{fig:GGA}(c)-(f)]. There are three elements 
contributing to these states, (\textit{i}) the impurity levels
(Mn $d$--orbitals), (\textit{ii}) the Mn-acceptor states, 
and (\textit{iii}) the TSSs. 
Figure~\ref{fig:DFT_DOS} shows the  
spin-resolved density of states (DOS) around the Fermi energy 
for the $p$--orbitals 
of Se atoms on the top and bottom surfaces and 
for the Mn $d$--orbitals, for $U$=$0$ and $U$=$4$~eV. \\
The calculated magnetic moment of the Mn atom on the surface, 
with spin orbit interaction (SOI), is 4.67~$\mu_B$ for $U$=$4$~eV, indicating 
that a substitutional Mn is close to its $+2$ valence state. 
Given the nominal $+3$ valence state of Bi in Bi$_2$Se$_3$, we conclude 
that the substitution of a Bi with a Mn introduces 
an acceptor (hole) state. Its 
wave function is localized primarily 
on the surroundings of the dopant and, to a lesser degree, on the dopant itself. 
For $U$=$0$ some of the Mn $d$--orbitals 
appear close to the Fermi level [Fig.~\ref{fig:DFT_DOS}(c)], 
in the same energy 
range as the Se $p$--orbitals, 
leading to their hybridization. 
Importantly, the top surface Se $p$--states 
around the Fermi level are visibly spin-polarized [Fig.~\ref{fig:DFT_DOS}(a)]. 
The $d$-hybridized Se $p$--orbitals above the Fermi level close to the Mn are   
the main contributors to the Mn-acceptor (hole) states. 
With increasing $U$, the majority Mn $d$--orbitals are pushed  
deeper into the valence band [Fig.~\ref{fig:DFT_DOS}(f)], 
decreasing the hybridization 
with Se $p$--orbitals on the top surface. 
 As a result, the Mn magnetic moment increases 
 by $\sim 7\%$ with respect to the $U=0$ value   
  and the spin-polarization of the top Se $p$--states 
   decreases [Fig.~\ref{fig:DFT_DOS}(d)].
We find a similar dependence of the magnetic moment on electronic correlations  
for  Mn in the bulk, namely, 4.25~$\mu_B$ 
  for $U$=$0$ and 4.52~$\mu_B$ for $U$=$4$~eV. 
These results suggest that the discrepancy between recent DFT calculations, 
reporting the values for the Mn 
magnetic moment ranging from 
4~$\mu_B$~\cite{niu2011mn,abdalla2013topological} to 4.58~$\mu_B$~\cite{henk2012complex,henk2012topological} 
 for Bi$_2$Se$_3$ and Bi$_2$Te$_3$, might be originating 
 from different treatment of electron interactions 
 on the impurity site.\\
The appearance of these unoccupied states above the Fermi level,
spatially localized around the Mn, 
is an indication of the acceptor level. 
These states occur in the same energy range 
$ E\in [0, 0.13]$~eV 
of the TSS of the top surface, and energetically are not far 
from its Dirac point. This is crucial for the opening 
of the gap. 
In contrast, in the range $E\in [0, 0.13]$~eV, the bottom surface 
states [Figs.~\ref{fig:DFT_DOS}(b) and (e)] are 
essentially the TSS, with negligible coupling to the impurity wavefunction. 
Their linear dispersion 
 is preserved and still detectable in 
 the bandstructure in Fig.~\ref{fig:GGA}(d). \\
To clarify the nature of the Mn-acceptor independently 
of the TSS, 
we perform calculations without SOI, which greatly simplifies 
the electronic structure around the Fermi energy.
\begin{figure}[htp]
\begin{minipage}[h]{1.0\linewidth}
\centering
\includegraphics[scale=0.16]{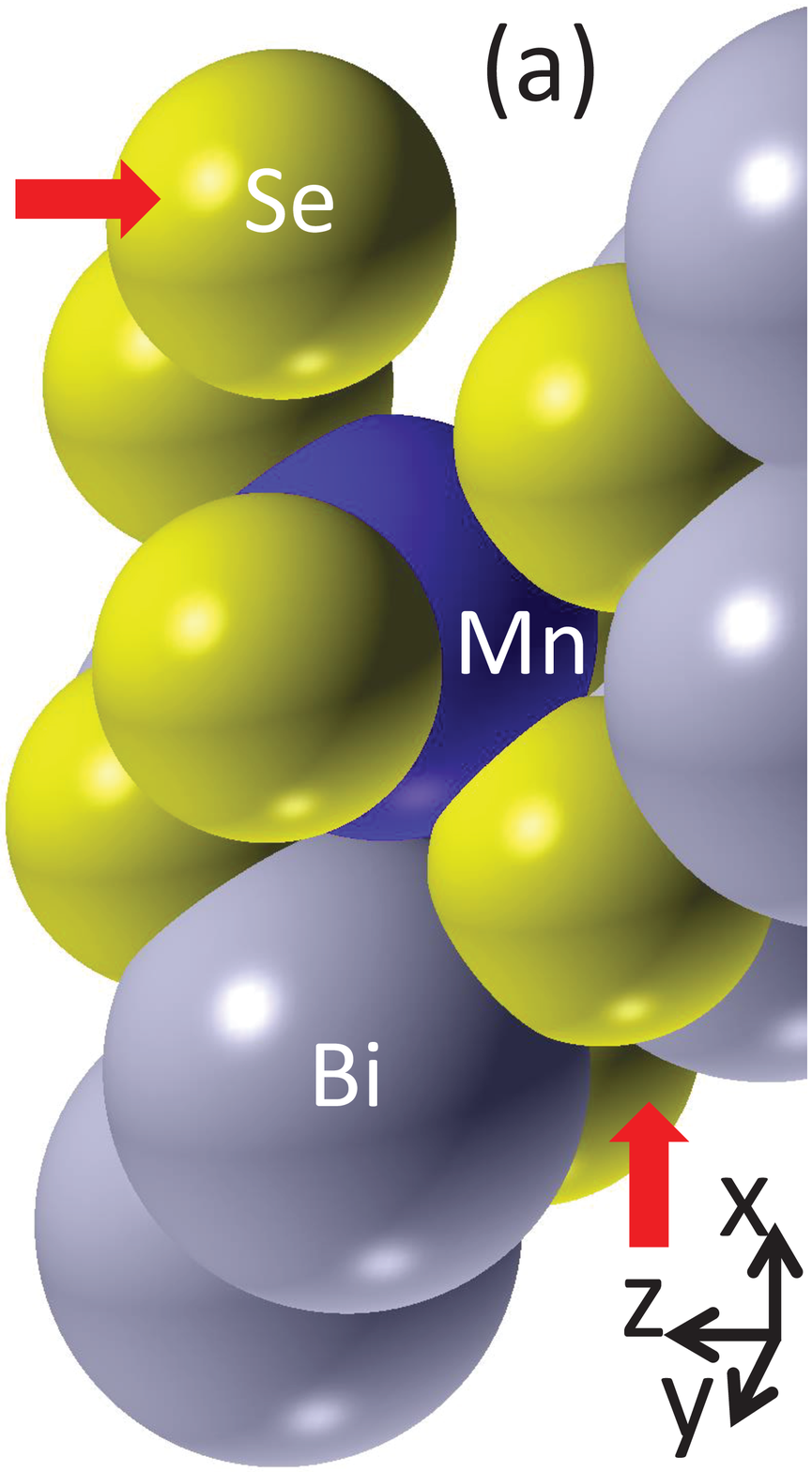}
\hspace{-1.50mm}
\includegraphics[scale=0.41]{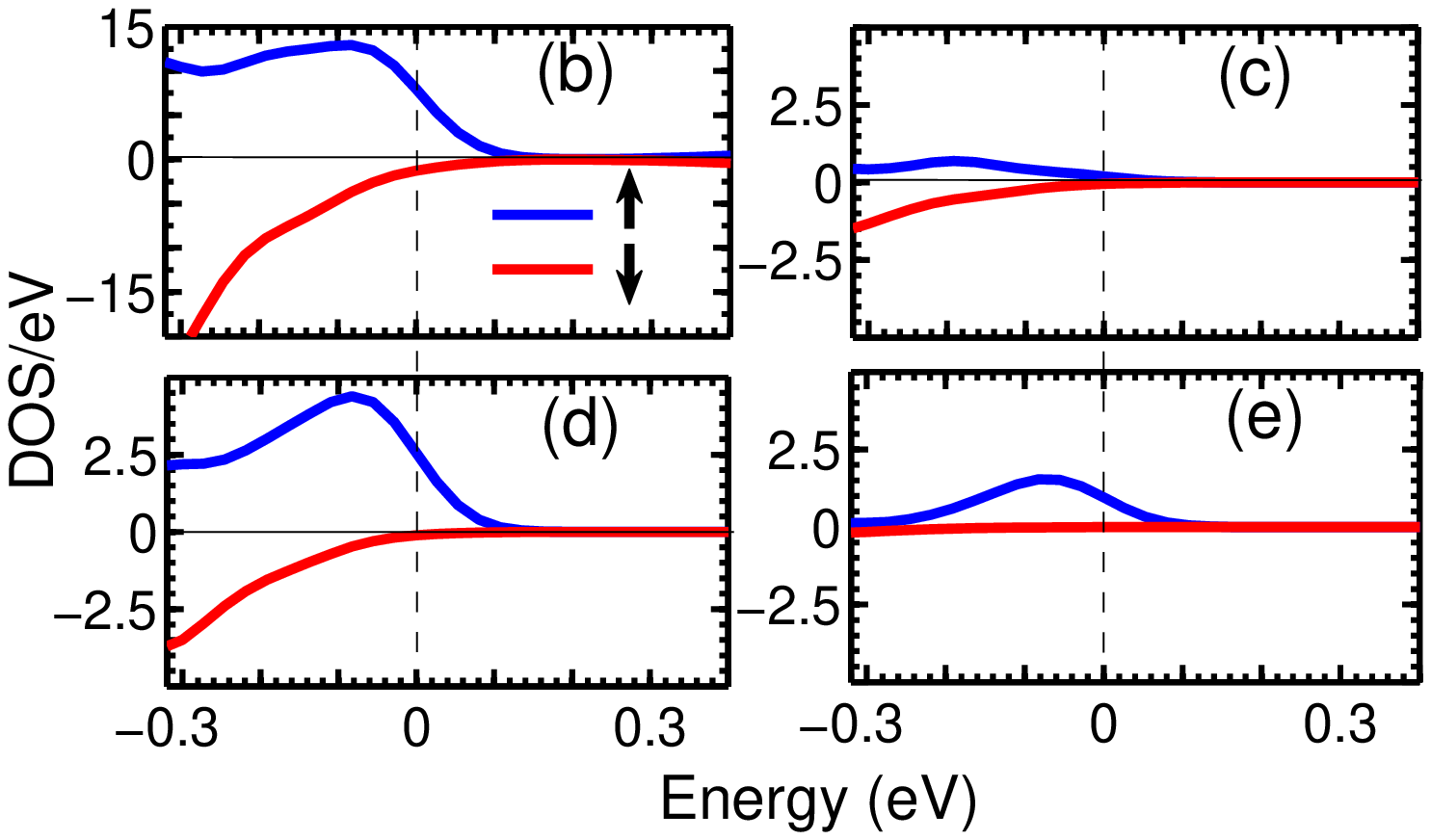}
\end{minipage}
		
\vspace{0.00mm}

\caption{(a) (111) Bi$_2$Se$_3$ 
surface with a Mn impurity in the second ML. 
DOS around the Fermi energy 
for Mn-doped Bi$_2$Se$_3$ slab for $U$=$0$ without SOI: 
(b) total DOS, 
(c) partial $p$--DOS for the two Se atoms above 
and below Mn, which are \textit{not} NN to Mn 
 [indicated by arrows in panel (a)],  
(d) partial $p$--DOS for the six NN Se atoms,  
(e) partial $d$--DOS for the Mn impurity. 
 Note that the total DOS in panel (b) includes 
 contributions from all atoms in the supercell and 
 from the interstitial region.}
\label{fig:nosoI}
\end{figure}
%
%
Figures~\ref{fig:nosoI}(b)-(e) show the total DOS and 
the partial DOS for the Mn impurity and surrounding Se atoms,   
calculated without SOI.  
We conclude that the peak in the total DOS (mainly majority spin) 
right below the Fermi level, is predominantly due to the 
Mn $d$--levels and the NN Se 
$p$--levels, with a very small but finite contribution 
from other Se atoms around the Mn [two Se atoms, 
which are not NN to Mn, are indicated by arrows in Fig.~\ref{fig:nosoI}(a)].  
The highly spin-polarized character of the $p$--states around 
the Fermi level is a consequence of  
the hybridization between the Mn $d-$orbitals and the NN Se $p-$orbitals. 
Our calculations with $U$=$4$~eV (not shown here) confirm 
this observation. Similar to 
the calculation with SOI (see Fig.~\ref{fig:DFT_DOS}), 
 for $U$=$4$~eV the Mn $d$-orbitals are more localized 
 and are pushed deeper into the valence band, 
which reduces the $p$--$d$ hybridization and decreases 
the polarization of NN Se $p$-states by a factor of two.\\ 
\begin{figure}
\begin{minipage}[h]{1.0\linewidth}
\centering
\includegraphics[scale=0.17]{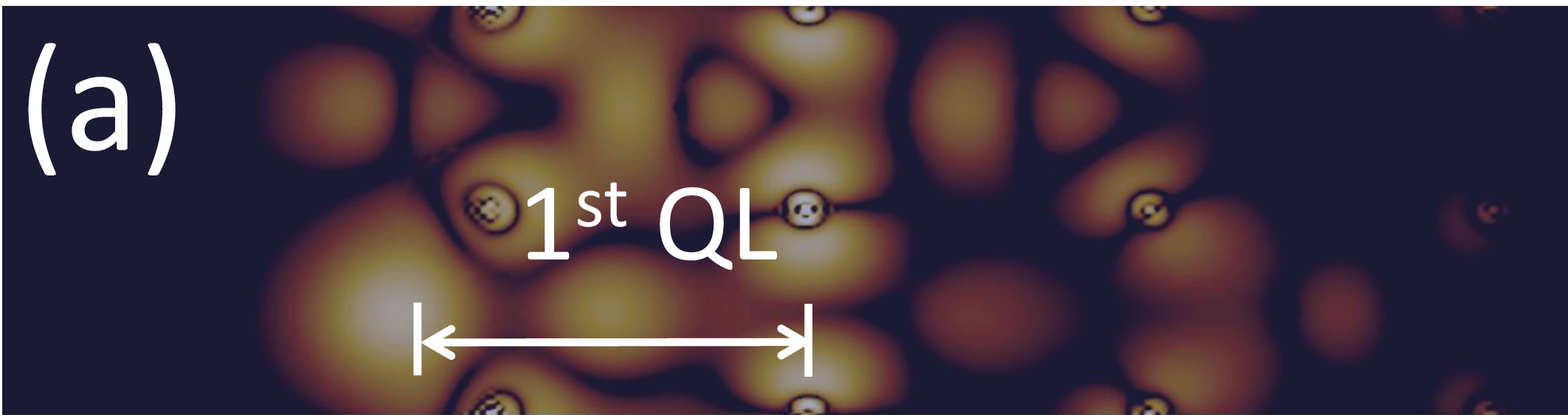}
\hspace{0.00mm}
\includegraphics[scale=0.137]{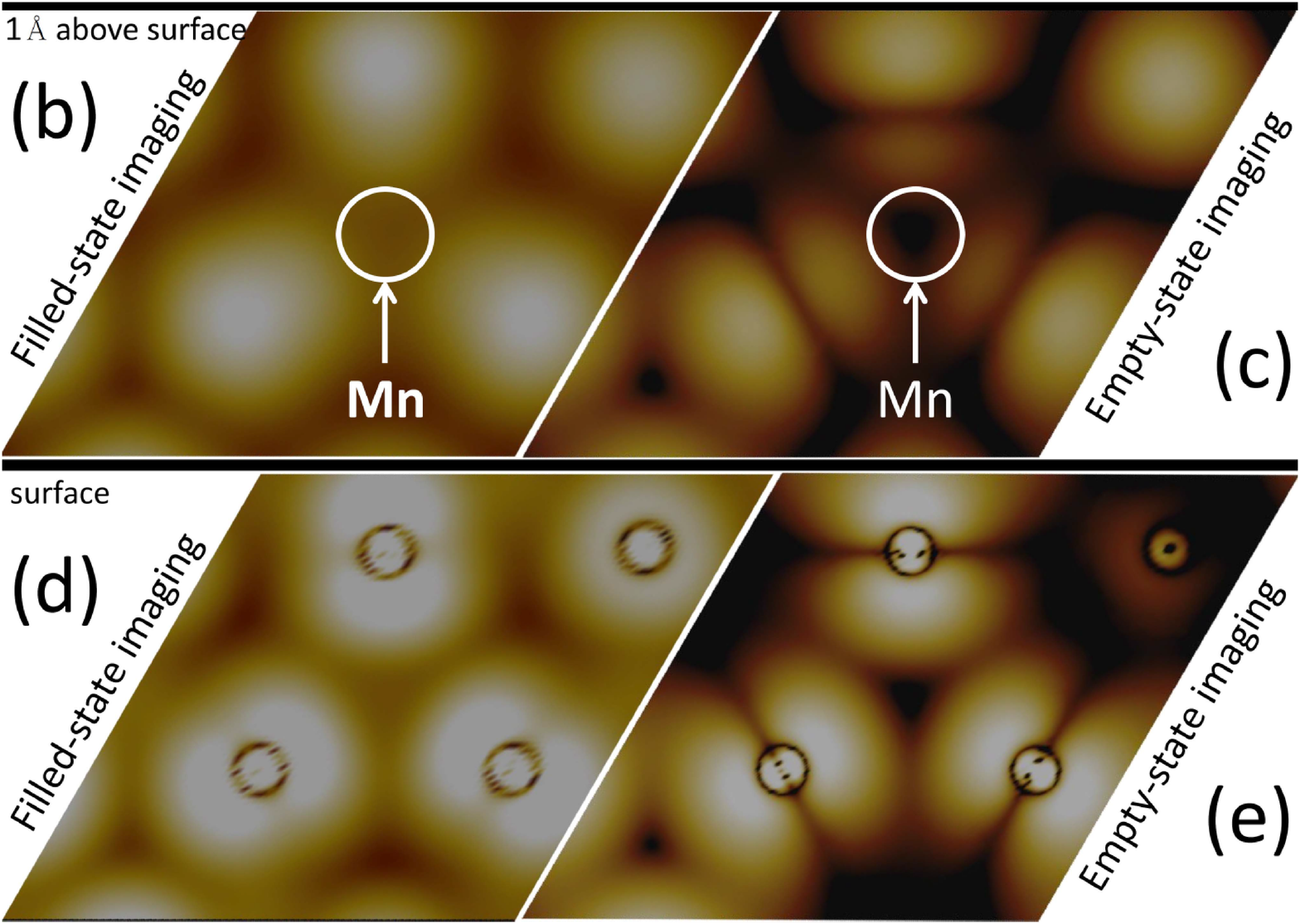}
\hspace{0.00mm}
\includegraphics[scale=0.17]{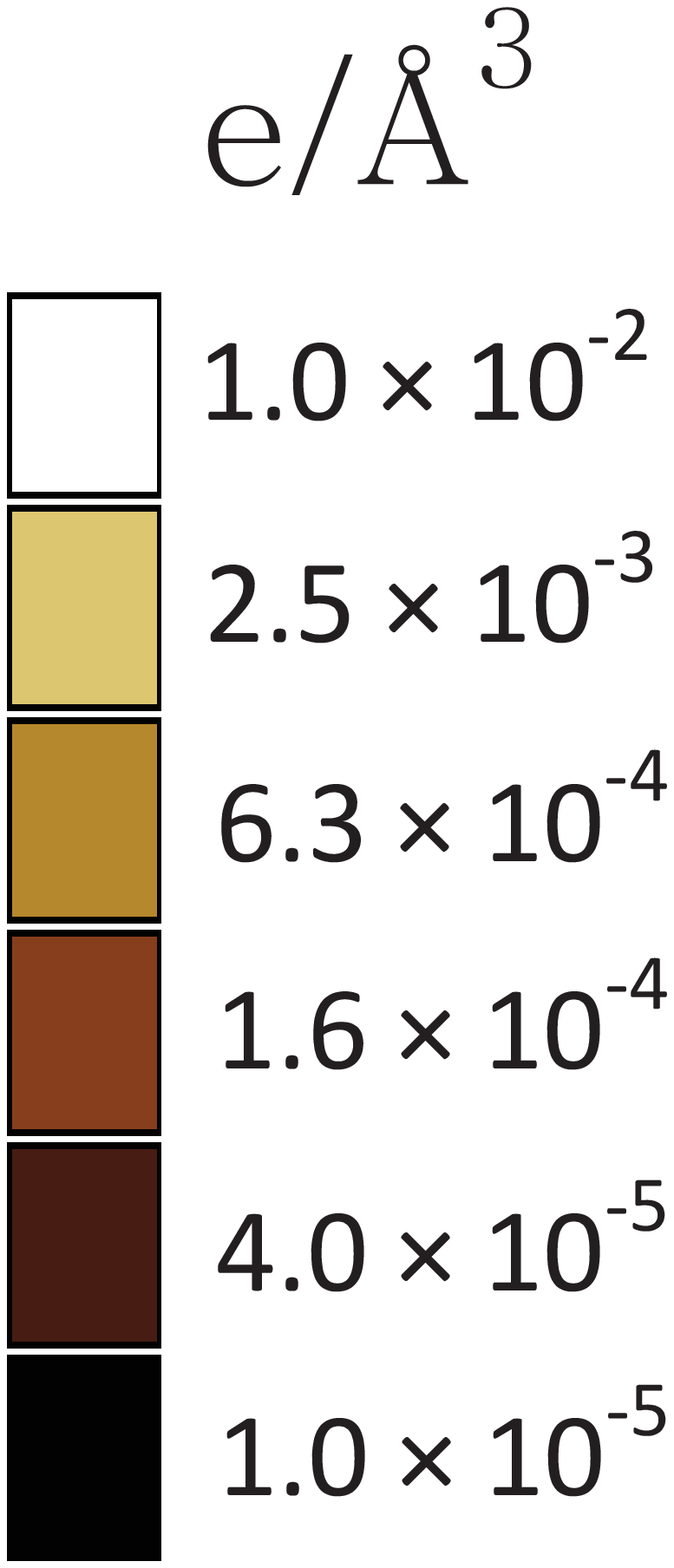}
\end{minipage}
		
\vspace{0.00mm}

\caption{LDOS 
 of Mn-doped Bi$_2$Se$_3$ integrated over 
the energy range $\pm0.13$~eV around the Fermi level. 
Positive (negative) energy 
 window correspond to empty-(filled-)state imaging. 
(a) Empty-state LDOS projected in the $xz$-plane, perpendicular to the (111) surface.  
LDOS projected on the the (111) surface: (b,c) at 1~$\AA$ above and   
(d,e) exactly on the surface. Left (right) panels are for   
 filled (empty) states. Note the logarithmic color-scale.}
\label{fig:chr-den}
\end{figure}
We now examine the spatial character of the Mn acceptor state, 
which is directly accessible by STM experiments.
Figures~\ref{fig:chr-den}(b)-(e) show simulated STM topographies 
of Mn-doped Bi$_2$Se$_3$ in the vicinity of the (111) 
surface. These images are obtained by plotting the 
electronic local density of states (LDOS) around the Mn, 
integrated in the energy 
window $[0, 0.13]$~eV (empty states) and $[-0.13, 0]$~eV (filled states) 
around the Fermi level for $U$=$0$. 
Figures~\ref{fig:chr-den}(c) and (e) clearly show that the acceptor state is 
predominately localized around the Mn and its three NN Se atoms.  
The state is composed of three $p$-like Se orbitals pointing to the Mn in the middle, visibly deformed by the hybridization.  
 It exhibits a characteristic triangular shape similar to the experimental STM topography observed at positive bias in
Mn-doped Bi$_2$Te$_3$~\cite{PhysRevB.81.195203}. 
The LDOS for filled states below the Fermi level [Figs.~\ref{fig:chr-den}(b) and (d)] are much less affected 
by the presence of the impurity.\\ 
The side view of the empty-state LDOS along the slab [Fig.~\ref{fig:chr-den}(a)] 
 confirms that the states in the energy range $[0, 0.13]$~eV are  
predominantly localized around Mn and its NN Se atoms, which is a signature of the Mn-acceptor. 
The figure also shows states that extend only within $\sim$1 QL from the surface, which is a typical  
  decay length of the TSS~\cite{ap_TI_prob}.
Clearly, the Mn-acceptor has a strong spatial overlap with the TSS of the top surface. 
Furthermore, as shown above, these two states are quasi-degenerate in the energy range $[0, 0.13]$ eV, 
and therefore they couple strongly.
It is precisely the quasi-resonant coupling of the spin-chiral TSS with the spin-polarized Mn-acceptor 
that ultimately opens a gap at the Dirac point. Strong support for this mechanism  is provided by the observation that the gap decreases 
from $5.5$~meV to $3.2$~meV when $U$ increases from 0 to 4~eV. Strong correlations at the impurity site decrease 
the Mn $d$-- and Se $p$--orbital hybridization, leading to a smaller spin-polarization of the acceptor. 
Isolated spin-polarized Mn $d$--levels 
are further away both in energy and in space from the TSS at the Dirac point.  
Therefore their spin-dependent potential alone is less effective in inducing the TRS breaking necessary to open a gap.\\
\section{Conclusions}
In conclusion, our calculations show, in agreement with 
experiments~\cite{PhysRevB.81.195203,APL_152103,phys_20083553}, that 
substitutional Mn impurities on Bi$_2$Se$_3$ surface 
introduce spin-polarized acceptor states, 
whose properties are similar to Mn-acceptors in GaAs.  
 The mechanism for 
the opening of a gap at the Dirac point is provided by 
the spatial overlap and the 
quasi-resonant coupling between the Mn-acceptor
and the TSS  
inside the bulk band gap of Bi$_2$Se$_3$. 
The signatures of this coupling can be detected in STM experiments, 
addressing specifically 
magnetic dopants on a 3D TI surface. 
The present study contributes to clarify the origin of surface-ferromagnetism 
in transition-metal-doped Bi-chalcogenide thin films.\\
Finally, we should mention that recent infrared optical experiments in Mn-doped 
Bi$_2$Te$_3$ thin films suggest that, despite the similarities to DMSs, carrier-independent 
mechanisms such as super-exchange~\cite{niu2011mn} and the aforementioned enhanced Van Vleck spin 
susceptibility~\cite{Yu02072010}, might also be relevant for establishing
the ferromagnetic state~\cite{PhysRevB.89.235308}. Specifically these experiments indicate that bulk charge 
carriers control the optical response but do not seem to play a significant role in 
mediating ferromagnetism. Note, however, that Mn-doped systems investigated in 
Ref.~\onlinecite{PhysRevB.89.235308} are always n-type rather than the expected p-type for substitutional Mn, 
with the Fermi energy always located in the Bi$_2$Te$_3$ conduction band. The reason of this 
fact is still unclear. In any case these systems are in a regime quite different from 
the one studied in the present paper, where the position of Fermi energy is 
characteristic of a p-type DMS. 
Further experimental and theoretical 
studies, addressing the role of bulk dopants and 
the position of the Fermi level in the bulk band gap,  	
are necessary to elucidate this point.\\
\section*{Acknowledgments}
This work was supported by the Faculty of Technology
at Linnaeus University, by the
Swedish Research Council under Grant Number: 621-2010-3761, 
and the NordForsk research network 080134 ``Nanospintronics: theory and
simulations".
Computational resources have
been provided by the Lunarc center for scientific and technical computing at
Lund University.
\bibliography{TI}

\begin{thebibliography}{37}%
\makeatletter
\providecommand \@ifxundefined [1]{%
 \@ifx{#1\undefined}
}%
\providecommand \@ifnum [1]{%
 \ifnum #1\expandafter \@firstoftwo
 \else \expandafter \@secondoftwo
 \fi
}%
\providecommand \@ifx [1]{%
 \ifx #1\expandafter \@firstoftwo
 \else \expandafter \@secondoftwo
 \fi
}%
\providecommand \natexlab [1]{#1}%
\providecommand \enquote  [1]{``#1''}%
\providecommand \bibnamefont  [1]{#1}%
\providecommand \bibfnamefont [1]{#1}%
\providecommand \citenamefont [1]{#1}%
\providecommand \href@noop [0]{\@secondoftwo}%
\providecommand \href [0]{\begingroup \@sanitize@url \@href}%
\providecommand \@href[1]{\@@startlink{#1}\@@href}%
\providecommand \@@href[1]{\endgroup#1\@@endlink}%
\providecommand \@sanitize@url [0]{\catcode `\\12\catcode `\$12\catcode
  `\&12\catcode `\#12\catcode `\^12\catcode `\_12\catcode `\%12\relax}%
\providecommand \@@startlink[1]{}%
\providecommand \@@endlink[0]{}%
\providecommand \url  [0]{\begingroup\@sanitize@url \@url }%
\providecommand \@url [1]{\endgroup\@href {#1}{\urlprefix }}%
\providecommand \urlprefix  [0]{URL }%
\providecommand \Eprint [0]{\href }%
\providecommand \doibase [0]{http://dx.doi.org/}%
\providecommand \selectlanguage [0]{\@gobble}%
\providecommand \bibinfo  [0]{\@secondoftwo}%
\providecommand \bibfield  [0]{\@secondoftwo}%
\providecommand \translation [1]{[#1]}%
\providecommand \BibitemOpen [0]{}%
\providecommand \bibitemStop [0]{}%
\providecommand \bibitemNoStop [0]{.\EOS\space}%
\providecommand \EOS [0]{\spacefactor3000\relax}%
\providecommand \BibitemShut  [1]{\csname bibitem#1\endcsname}%
\let\auto@bib@innerbib\@empty
\bibitem [{\citenamefont {Hasan}\ and\ \citenamefont {Kane}(2010)}]{Hasan}%
  \BibitemOpen
  \bibfield  {author} {\bibinfo {author} {\bibfnamefont {M.~Z.}\ \bibnamefont
  {Hasan}}\ and\ \bibinfo {author} {\bibfnamefont {C.~L.}\ \bibnamefont
  {Kane}},\ }\href {\doibase 10.1103/RevModPhys.82.3045} {\bibfield  {journal}
  {\bibinfo  {journal} {Rev. Mod. Phys.}\ }\textbf {\bibinfo {volume} {82}},\
  \bibinfo {pages} {3045} (\bibinfo {year} {2010})}\BibitemShut {NoStop}%
\bibitem [{\citenamefont {Qi}\ and\ \citenamefont {Zhang}(2011)}]{XLQi}%
  \BibitemOpen
  \bibfield  {author} {\bibinfo {author} {\bibfnamefont {X.-L.}\ \bibnamefont
  {Qi}}\ and\ \bibinfo {author} {\bibfnamefont {S.-C.}\ \bibnamefont {Zhang}},\
  }\href {\doibase 10.1103/RevModPhys.83.1057} {\bibfield  {journal} {\bibinfo
  {journal} {Rev. Mod. Phys.}\ }\textbf {\bibinfo {volume} {83}},\ \bibinfo
  {pages} {1057} (\bibinfo {year} {2011})}\BibitemShut {NoStop}%
\bibitem [{\citenamefont {Zhang}\ \emph {et~al.}(2009)\citenamefont {Zhang},
  \citenamefont {Liu}, \citenamefont {Qi}, \citenamefont {Dai}, \citenamefont
  {Fang},\ and\ \citenamefont {Zhang}}]{zhang2009topological}%
  \BibitemOpen
  \bibfield  {author} {\bibinfo {author} {\bibfnamefont {H.}~\bibnamefont
  {Zhang}}, \bibinfo {author} {\bibfnamefont {C.-X.}\ \bibnamefont {Liu}},
  \bibinfo {author} {\bibfnamefont {X.-L.}\ \bibnamefont {Qi}}, \bibinfo
  {author} {\bibfnamefont {X.}~\bibnamefont {Dai}}, \bibinfo {author}
  {\bibfnamefont {Z.}~\bibnamefont {Fang}}, \ and\ \bibinfo {author}
  {\bibfnamefont {S.-C.}\ \bibnamefont {Zhang}},\ }\href@noop {} {\bibfield
  {journal} {\bibinfo  {journal} {Nature Physics}\ }\textbf {\bibinfo {volume}
  {5}},\ \bibinfo {pages} {438} (\bibinfo {year} {2009})}\BibitemShut {NoStop}%
\bibitem [{\citenamefont {Yu}\ \emph {et~al.}(2010)\citenamefont {Yu},
  \citenamefont {Zhang}, \citenamefont {Zhang}, \citenamefont {Zhang},
  \citenamefont {Dai},\ and\ \citenamefont {Fang}}]{Yu02072010}%
  \BibitemOpen
  \bibfield  {author} {\bibinfo {author} {\bibfnamefont {R.}~\bibnamefont
  {Yu}}, \bibinfo {author} {\bibfnamefont {W.}~\bibnamefont {Zhang}}, \bibinfo
  {author} {\bibfnamefont {H.-J.}\ \bibnamefont {Zhang}}, \bibinfo {author}
  {\bibfnamefont {S.-C.}\ \bibnamefont {Zhang}}, \bibinfo {author}
  {\bibfnamefont {X.}~\bibnamefont {Dai}}, \ and\ \bibinfo {author}
  {\bibfnamefont {Z.}~\bibnamefont {Fang}},\ }\href {\doibase
  10.1126/science.1187485} {\bibfield  {journal} {\bibinfo  {journal}
  {Science}\ }\textbf {\bibinfo {volume} {329}},\ \bibinfo {pages} {61}
  (\bibinfo {year} {2010})}\BibitemShut {NoStop}%
\bibitem [{\citenamefont {Chang}\ \emph {et~al.}(2013)\citenamefont {Chang},
  \citenamefont {Zhang}, \citenamefont {Feng}, \citenamefont {Shen},
  \citenamefont {Zhang}, \citenamefont {Guo}, \citenamefont {Li}, \citenamefont
  {Ou}, \citenamefont {Wei}, \citenamefont {Wang}, \citenamefont {Ji},
  \citenamefont {Feng}, \citenamefont {Ji}, \citenamefont {Chen}, \citenamefont
  {Jia}, \citenamefont {Dai}, \citenamefont {Fang}, \citenamefont {Zhang},
  \citenamefont {He}, \citenamefont {Wang}, \citenamefont {Lu}, \citenamefont
  {Ma},\ and\ \citenamefont {Xue}}]{Chang12042013}%
  \BibitemOpen
  \bibfield  {author} {\bibinfo {author} {\bibfnamefont {C.-Z.}\ \bibnamefont
  {Chang}}, \bibinfo {author} {\bibfnamefont {J.}~\bibnamefont {Zhang}},
  \bibinfo {author} {\bibfnamefont {X.}~\bibnamefont {Feng}}, \bibinfo {author}
  {\bibfnamefont {J.}~\bibnamefont {Shen}}, \bibinfo {author} {\bibfnamefont
  {Z.}~\bibnamefont {Zhang}}, \bibinfo {author} {\bibfnamefont
  {M.}~\bibnamefont {Guo}}, \bibinfo {author} {\bibfnamefont {K.}~\bibnamefont
  {Li}}, \bibinfo {author} {\bibfnamefont {Y.}~\bibnamefont {Ou}}, \bibinfo
  {author} {\bibfnamefont {P.}~\bibnamefont {Wei}}, \bibinfo {author}
  {\bibfnamefont {L.-L.}\ \bibnamefont {Wang}}, \bibinfo {author}
  {\bibfnamefont {Z.-Q.}\ \bibnamefont {Ji}}, \bibinfo {author} {\bibfnamefont
  {Y.}~\bibnamefont {Feng}}, \bibinfo {author} {\bibfnamefont {S.}~\bibnamefont
  {Ji}}, \bibinfo {author} {\bibfnamefont {X.}~\bibnamefont {Chen}}, \bibinfo
  {author} {\bibfnamefont {J.}~\bibnamefont {Jia}}, \bibinfo {author}
  {\bibfnamefont {X.}~\bibnamefont {Dai}}, \bibinfo {author} {\bibfnamefont
  {Z.}~\bibnamefont {Fang}}, \bibinfo {author} {\bibfnamefont {S.-C.}\
  \bibnamefont {Zhang}}, \bibinfo {author} {\bibfnamefont {K.}~\bibnamefont
  {He}}, \bibinfo {author} {\bibfnamefont {Y.}~\bibnamefont {Wang}}, \bibinfo
  {author} {\bibfnamefont {L.}~\bibnamefont {Lu}}, \bibinfo {author}
  {\bibfnamefont {X.-C.}\ \bibnamefont {Ma}}, \ and\ \bibinfo {author}
  {\bibfnamefont {Q.-K.}\ \bibnamefont {Xue}},\ }\href {\doibase
  10.1126/science.1234414} {\bibfield  {journal} {\bibinfo  {journal}
  {Science}\ }\textbf {\bibinfo {volume} {340}},\ \bibinfo {pages} {167}
  (\bibinfo {year} {2013})}\BibitemShut {NoStop}%
\bibitem [{\citenamefont {Qi}\ \emph {et~al.}(2008)\citenamefont {Qi},
  \citenamefont {Hughes},\ and\ \citenamefont {Zhang}}]{PhysRevB.78.195424}%
  \BibitemOpen
  \bibfield  {author} {\bibinfo {author} {\bibfnamefont {X.-L.}\ \bibnamefont
  {Qi}}, \bibinfo {author} {\bibfnamefont {T.~L.}\ \bibnamefont {Hughes}}, \
  and\ \bibinfo {author} {\bibfnamefont {S.-C.}\ \bibnamefont {Zhang}},\ }\href
  {\doibase 10.1103/PhysRevB.78.195424} {\bibfield  {journal} {\bibinfo
  {journal} {Phys. Rev. B}\ }\textbf {\bibinfo {volume} {78}},\ \bibinfo
  {pages} {195424} (\bibinfo {year} {2008})}\BibitemShut {NoStop}%
\bibitem [{\citenamefont {Liu}\ \emph {et~al.}(2009)\citenamefont {Liu},
  \citenamefont {Liu}, \citenamefont {Xu}, \citenamefont {Qi},\ and\
  \citenamefont {Zhang}}]{PhysRevLett.102.156603}%
  \BibitemOpen
  \bibfield  {author} {\bibinfo {author} {\bibfnamefont {Q.}~\bibnamefont
  {Liu}}, \bibinfo {author} {\bibfnamefont {C.-X.}\ \bibnamefont {Liu}},
  \bibinfo {author} {\bibfnamefont {C.}~\bibnamefont {Xu}}, \bibinfo {author}
  {\bibfnamefont {X.-L.}\ \bibnamefont {Qi}}, \ and\ \bibinfo {author}
  {\bibfnamefont {S.-C.}\ \bibnamefont {Zhang}},\ }\href {\doibase
  10.1103/PhysRevLett.102.156603} {\bibfield  {journal} {\bibinfo  {journal}
  {Phys. Rev. Lett.}\ }\textbf {\bibinfo {volume} {102}},\ \bibinfo {pages}
  {156603} (\bibinfo {year} {2009})}\BibitemShut {NoStop}%
\bibitem [{\citenamefont {Checkelsky}\ \emph {et~al.}(2012)\citenamefont
  {Checkelsky}, \citenamefont {Ye}, \citenamefont {Onose},\ and\ \citenamefont
  {Tokura}}]{Checkelsky}%
  \BibitemOpen
  \bibfield  {author} {\bibinfo {author} {\bibfnamefont {J.~G.}\ \bibnamefont
  {Checkelsky}}, \bibinfo {author} {\bibfnamefont {J.}~\bibnamefont {Ye}},
  \bibinfo {author} {\bibfnamefont {Y.}~\bibnamefont {Onose}}, \ and\ \bibinfo
  {author} {\bibfnamefont {Y.}~\bibnamefont {Tokura}},\ }\href {\doibase
  doi:10.1038/nphys2388} {\bibfield  {journal} {\bibinfo  {journal} {Nature
  Physics}\ }\textbf {\bibinfo {volume} {8}},\ \bibinfo {pages} {729} (\bibinfo
  {year} {2012})}\BibitemShut {NoStop}%
\bibitem [{\citenamefont {Niu}\ \emph {et~al.}(2011)\citenamefont {Niu},
  \citenamefont {Dai}, \citenamefont {Guo}, \citenamefont {Wei}, \citenamefont
  {Ma},\ and\ \citenamefont {Huang}}]{niu2011mn}%
  \BibitemOpen
  \bibfield  {author} {\bibinfo {author} {\bibfnamefont {C.}~\bibnamefont
  {Niu}}, \bibinfo {author} {\bibfnamefont {Y.}~\bibnamefont {Dai}}, \bibinfo
  {author} {\bibfnamefont {M.}~\bibnamefont {Guo}}, \bibinfo {author}
  {\bibfnamefont {W.}~\bibnamefont {Wei}}, \bibinfo {author} {\bibfnamefont
  {Y.}~\bibnamefont {Ma}}, \ and\ \bibinfo {author} {\bibfnamefont
  {B.}~\bibnamefont {Huang}},\ }\href@noop {} {\bibfield  {journal} {\bibinfo
  {journal} {Appl. Phys. Lett.}\ }\textbf {\bibinfo {volume} {98}},\ \bibinfo
  {pages} {252502} (\bibinfo {year} {2011})}\BibitemShut {NoStop}%
\bibitem [{\citenamefont {Henk}\ \emph
  {et~al.}(2012{\natexlab{a}})\citenamefont {Henk}, \citenamefont {Flieger},
  \citenamefont {Maznichenko}, \citenamefont {Mertig}, \citenamefont {Ernst},
  \citenamefont {Eremeev},\ and\ \citenamefont
  {Chulkov}}]{henk2012topological}%
  \BibitemOpen
  \bibfield  {author} {\bibinfo {author} {\bibfnamefont {J.}~\bibnamefont
  {Henk}}, \bibinfo {author} {\bibfnamefont {M.}~\bibnamefont {Flieger}},
  \bibinfo {author} {\bibfnamefont {I.~V.}\ \bibnamefont {Maznichenko}},
  \bibinfo {author} {\bibfnamefont {I.}~\bibnamefont {Mertig}}, \bibinfo
  {author} {\bibfnamefont {A.}~\bibnamefont {Ernst}}, \bibinfo {author}
  {\bibfnamefont {S.~V.}\ \bibnamefont {Eremeev}}, \ and\ \bibinfo {author}
  {\bibfnamefont {E.~V.}\ \bibnamefont {Chulkov}},\ }\href {\doibase
  10.1103/PhysRevLett.109.076801} {\bibfield  {journal} {\bibinfo  {journal}
  {Phys. Rev. Lett.}\ }\textbf {\bibinfo {volume} {109}},\ \bibinfo {pages}
  {076801} (\bibinfo {year} {2012}{\natexlab{a}})}\BibitemShut {NoStop}%
\bibitem [{\citenamefont {Abdalla}\ \emph {et~al.}(2013)\citenamefont
  {Abdalla}, \citenamefont {Seixas}, \citenamefont {Schmidt}, \citenamefont
  {Miwa},\ and\ \citenamefont {Fazzio}}]{abdalla2013topological}%
  \BibitemOpen
  \bibfield  {author} {\bibinfo {author} {\bibfnamefont {L.~B.}\ \bibnamefont
  {Abdalla}}, \bibinfo {author} {\bibfnamefont {L.}~\bibnamefont {Seixas}},
  \bibinfo {author} {\bibfnamefont {T.~M.}\ \bibnamefont {Schmidt}}, \bibinfo
  {author} {\bibfnamefont {R.~H.}\ \bibnamefont {Miwa}}, \ and\ \bibinfo
  {author} {\bibfnamefont {A.}~\bibnamefont {Fazzio}},\ }\href {\doibase
  10.1103/PhysRevB.88.045312} {\bibfield  {journal} {\bibinfo  {journal} {Phys.
  Rev. B}\ }\textbf {\bibinfo {volume} {88}},\ \bibinfo {pages} {045312}
  (\bibinfo {year} {2013})}\BibitemShut {NoStop}%
\bibitem [{\citenamefont {Li}\ \emph {et~al.}(2014)\citenamefont {Li},
  \citenamefont {Zou}, \citenamefont {Li},\ and\ \citenamefont
  {Zhou}}]{Yuanchang}%
  \BibitemOpen
  \bibfield  {author} {\bibinfo {author} {\bibfnamefont {Y.}~\bibnamefont
  {Li}}, \bibinfo {author} {\bibfnamefont {X.}~\bibnamefont {Zou}}, \bibinfo
  {author} {\bibfnamefont {J.}~\bibnamefont {Li}}, \ and\ \bibinfo {author}
  {\bibfnamefont {G.}~\bibnamefont {Zhou}},\ }\href {\doibase
  http://dx.doi.org/10.1063/1.4869146} {\bibfield  {journal} {\bibinfo
  {journal} {The Journal of Chemical Physics}\ }\textbf {\bibinfo {volume}
  {140}},\ \bibinfo {eid} {124704} (\bibinfo {year} {2014})}\BibitemShut
  {NoStop}%
\bibitem [{\citenamefont {Chen}\ \emph {et~al.}(2010)\citenamefont {Chen},
  \citenamefont {Chu}, \citenamefont {Analytis}, \citenamefont {Liu},
  \citenamefont {Igarashi}, \citenamefont {Kuo}, \citenamefont {Qi},
  \citenamefont {Mo}, \citenamefont {Moore}, \citenamefont {Lu}, \citenamefont
  {Hashimoto}, \citenamefont {Sasagawa}, \citenamefont {Zhang}, \citenamefont
  {Fisher}, \citenamefont {Hussain},\ and\ \citenamefont
  {Shen}}]{Chen06082010}%
  \BibitemOpen
  \bibfield  {author} {\bibinfo {author} {\bibfnamefont {Y.~L.}\ \bibnamefont
  {Chen}}, \bibinfo {author} {\bibfnamefont {J.-H.}\ \bibnamefont {Chu}},
  \bibinfo {author} {\bibfnamefont {J.~G.}\ \bibnamefont {Analytis}}, \bibinfo
  {author} {\bibfnamefont {Z.~K.}\ \bibnamefont {Liu}}, \bibinfo {author}
  {\bibfnamefont {K.}~\bibnamefont {Igarashi}}, \bibinfo {author}
  {\bibfnamefont {H.-H.}\ \bibnamefont {Kuo}}, \bibinfo {author} {\bibfnamefont
  {X.~L.}\ \bibnamefont {Qi}}, \bibinfo {author} {\bibfnamefont {S.~K.}\
  \bibnamefont {Mo}}, \bibinfo {author} {\bibfnamefont {R.~G.}\ \bibnamefont
  {Moore}}, \bibinfo {author} {\bibfnamefont {D.~H.}\ \bibnamefont {Lu}},
  \bibinfo {author} {\bibfnamefont {M.}~\bibnamefont {Hashimoto}}, \bibinfo
  {author} {\bibfnamefont {T.}~\bibnamefont {Sasagawa}}, \bibinfo {author}
  {\bibfnamefont {S.~C.}\ \bibnamefont {Zhang}}, \bibinfo {author}
  {\bibfnamefont {I.~R.}\ \bibnamefont {Fisher}}, \bibinfo {author}
  {\bibfnamefont {Z.}~\bibnamefont {Hussain}}, \ and\ \bibinfo {author}
  {\bibfnamefont {Z.~X.}\ \bibnamefont {Shen}},\ }\href {\doibase
  10.1126/science.1189924} {\bibfield  {journal} {\bibinfo  {journal}
  {Science}\ }\textbf {\bibinfo {volume} {329}},\ \bibinfo {pages} {659}
  (\bibinfo {year} {2010})}\BibitemShut {NoStop}%
\bibitem [{\citenamefont {Scholz}\ \emph {et~al.}(2012)\citenamefont {Scholz},
  \citenamefont {S\'anchez-Barriga}, \citenamefont {Marchenko}, \citenamefont
  {Varykhalov}, \citenamefont {Volykhov}, \citenamefont {Yashina},\ and\
  \citenamefont {Rader}}]{PhysRevLett.108.256810}%
  \BibitemOpen
  \bibfield  {author} {\bibinfo {author} {\bibfnamefont {M.~R.}\ \bibnamefont
  {Scholz}}, \bibinfo {author} {\bibfnamefont {J.}~\bibnamefont
  {S\'anchez-Barriga}}, \bibinfo {author} {\bibfnamefont {D.}~\bibnamefont
  {Marchenko}}, \bibinfo {author} {\bibfnamefont {A.}~\bibnamefont
  {Varykhalov}}, \bibinfo {author} {\bibfnamefont {A.}~\bibnamefont
  {Volykhov}}, \bibinfo {author} {\bibfnamefont {L.~V.}\ \bibnamefont
  {Yashina}}, \ and\ \bibinfo {author} {\bibfnamefont {O.}~\bibnamefont
  {Rader}},\ }\href {\doibase 10.1103/PhysRevLett.108.256810} {\bibfield
  {journal} {\bibinfo  {journal} {Phys. Rev. Lett.}\ }\textbf {\bibinfo
  {volume} {108}},\ \bibinfo {pages} {256810} (\bibinfo {year}
  {2012})}\BibitemShut {NoStop}%
\bibitem [{\citenamefont {Valla}\ \emph {et~al.}(2012)\citenamefont {Valla},
  \citenamefont {Pan}, \citenamefont {Gardner}, \citenamefont {Lee},\ and\
  \citenamefont {Chu}}]{PhysRevLett.108.117601}%
  \BibitemOpen
  \bibfield  {author} {\bibinfo {author} {\bibfnamefont {T.}~\bibnamefont
  {Valla}}, \bibinfo {author} {\bibfnamefont {Z.-H.}\ \bibnamefont {Pan}},
  \bibinfo {author} {\bibfnamefont {D.}~\bibnamefont {Gardner}}, \bibinfo
  {author} {\bibfnamefont {Y.~S.}\ \bibnamefont {Lee}}, \ and\ \bibinfo
  {author} {\bibfnamefont {S.}~\bibnamefont {Chu}},\ }\href {\doibase
  10.1103/PhysRevLett.108.117601} {\bibfield  {journal} {\bibinfo  {journal}
  {Phys. Rev. Lett.}\ }\textbf {\bibinfo {volume} {108}},\ \bibinfo {pages}
  {117601} (\bibinfo {year} {2012})}\BibitemShut {NoStop}%
\bibitem [{\citenamefont {Honolka}\ \emph {et~al.}(2012)\citenamefont
  {Honolka}, \citenamefont {Khajetoorians}, \citenamefont {Sessi},
  \citenamefont {Wehling}, \citenamefont {Stepanow}, \citenamefont {Mi},
  \citenamefont {Iversen}, \citenamefont {Schlenk}, \citenamefont {Wiebe},
  \citenamefont {Brookes}, \citenamefont {Lichtenstein}, \citenamefont
  {Hofmann}, \citenamefont {Kern},\ and\ \citenamefont
  {Wiesendanger}}]{honolka2012plane}%
  \BibitemOpen
  \bibfield  {author} {\bibinfo {author} {\bibfnamefont {J.}~\bibnamefont
  {Honolka}}, \bibinfo {author} {\bibfnamefont {A.~A.}\ \bibnamefont
  {Khajetoorians}}, \bibinfo {author} {\bibfnamefont {V.}~\bibnamefont
  {Sessi}}, \bibinfo {author} {\bibfnamefont {T.~O.}\ \bibnamefont {Wehling}},
  \bibinfo {author} {\bibfnamefont {S.}~\bibnamefont {Stepanow}}, \bibinfo
  {author} {\bibfnamefont {J.-L.}\ \bibnamefont {Mi}}, \bibinfo {author}
  {\bibfnamefont {B.~B.}\ \bibnamefont {Iversen}}, \bibinfo {author}
  {\bibfnamefont {T.}~\bibnamefont {Schlenk}}, \bibinfo {author} {\bibfnamefont
  {J.}~\bibnamefont {Wiebe}}, \bibinfo {author} {\bibfnamefont {N.~B.}\
  \bibnamefont {Brookes}}, \bibinfo {author} {\bibfnamefont {A.~I.}\
  \bibnamefont {Lichtenstein}}, \bibinfo {author} {\bibfnamefont
  {P.}~\bibnamefont {Hofmann}}, \bibinfo {author} {\bibfnamefont
  {K.}~\bibnamefont {Kern}}, \ and\ \bibinfo {author} {\bibfnamefont
  {R.}~\bibnamefont {Wiesendanger}},\ }\href {\doibase
  10.1103/PhysRevLett.108.256811} {\bibfield  {journal} {\bibinfo  {journal}
  {Phys. Rev. Lett.}\ }\textbf {\bibinfo {volume} {108}},\ \bibinfo {pages}
  {256811} (\bibinfo {year} {2012})}\BibitemShut {NoStop}%
\bibitem [{\citenamefont {Schmidt}\ \emph {et~al.}(2013)\citenamefont
  {Schmidt}, \citenamefont {Miwa},\ and\ \citenamefont
  {Fazzio}}]{0953-8984-25-44-445003}%
  \BibitemOpen
  \bibfield  {author} {\bibinfo {author} {\bibfnamefont {T.~M.}\ \bibnamefont
  {Schmidt}}, \bibinfo {author} {\bibfnamefont {R.~H.}\ \bibnamefont {Miwa}}, \
  and\ \bibinfo {author} {\bibfnamefont {A.}~\bibnamefont {Fazzio}},\ }\href
  {http://stacks.iop.org/0953-8984/25/i=44/a=445003} {\bibfield  {journal}
  {\bibinfo  {journal} {Journal of Physics: Condensed Matter}\ }\textbf
  {\bibinfo {volume} {25}},\ \bibinfo {pages} {445003} (\bibinfo {year}
  {2013})}\BibitemShut {NoStop}%
\bibitem [{\citenamefont {Wray}\ \emph {et~al.}(2011)\citenamefont {Wray},
  \citenamefont {Xu}, \citenamefont {Xia}, \citenamefont {Hsieh}, \citenamefont
  {Fedorov}, \citenamefont {San~Hor}, \citenamefont {Cava}, \citenamefont
  {Bansil}, \citenamefont {Lin},\ and\ \citenamefont {Hasan}}]{Wray2010}%
  \BibitemOpen
  \bibfield  {author} {\bibinfo {author} {\bibfnamefont {L.~A.}\ \bibnamefont
  {Wray}}, \bibinfo {author} {\bibfnamefont {S.-Y.}\ \bibnamefont {Xu}},
  \bibinfo {author} {\bibfnamefont {Y.}~\bibnamefont {Xia}}, \bibinfo {author}
  {\bibfnamefont {D.}~\bibnamefont {Hsieh}}, \bibinfo {author} {\bibfnamefont
  {A.~V.}\ \bibnamefont {Fedorov}}, \bibinfo {author} {\bibfnamefont
  {Y.}~\bibnamefont {San~Hor}}, \bibinfo {author} {\bibfnamefont {R.~J.}\
  \bibnamefont {Cava}}, \bibinfo {author} {\bibfnamefont {A.}~\bibnamefont
  {Bansil}}, \bibinfo {author} {\bibfnamefont {H.}~\bibnamefont {Lin}}, \ and\
  \bibinfo {author} {\bibfnamefont {M.~Z.}\ \bibnamefont {Hasan}},\ }\href@noop
  {} {\bibfield  {journal} {\bibinfo  {journal} {Nature Physics}\ }\textbf
  {\bibinfo {volume} {7}},\ \bibinfo {pages} {32} (\bibinfo {year}
  {2011})}\BibitemShut {NoStop}%
\bibitem [{\citenamefont {Hor}\ \emph {et~al.}(2010)\citenamefont {Hor},
  \citenamefont {Roushan}, \citenamefont {Beidenkopf}, \citenamefont {Seo},
  \citenamefont {Qu}, \citenamefont {Checkelsky}, \citenamefont {Wray},
  \citenamefont {Hsieh}, \citenamefont {Xia}, \citenamefont {Xu}, \citenamefont
  {Qian}, \citenamefont {Hasan}, \citenamefont {Ong}, \citenamefont {Yazdani},\
  and\ \citenamefont {Cava}}]{PhysRevB.81.195203}%
  \BibitemOpen
  \bibfield  {author} {\bibinfo {author} {\bibfnamefont {Y.~S.}\ \bibnamefont
  {Hor}}, \bibinfo {author} {\bibfnamefont {P.}~\bibnamefont {Roushan}},
  \bibinfo {author} {\bibfnamefont {H.}~\bibnamefont {Beidenkopf}}, \bibinfo
  {author} {\bibfnamefont {J.}~\bibnamefont {Seo}}, \bibinfo {author}
  {\bibfnamefont {D.}~\bibnamefont {Qu}}, \bibinfo {author} {\bibfnamefont
  {J.~G.}\ \bibnamefont {Checkelsky}}, \bibinfo {author} {\bibfnamefont
  {L.~A.}\ \bibnamefont {Wray}}, \bibinfo {author} {\bibfnamefont
  {D.}~\bibnamefont {Hsieh}}, \bibinfo {author} {\bibfnamefont
  {Y.}~\bibnamefont {Xia}}, \bibinfo {author} {\bibfnamefont {S.-Y.}\
  \bibnamefont {Xu}}, \bibinfo {author} {\bibfnamefont {D.}~\bibnamefont
  {Qian}}, \bibinfo {author} {\bibfnamefont {M.~Z.}\ \bibnamefont {Hasan}},
  \bibinfo {author} {\bibfnamefont {N.~P.}\ \bibnamefont {Ong}}, \bibinfo
  {author} {\bibfnamefont {A.}~\bibnamefont {Yazdani}}, \ and\ \bibinfo
  {author} {\bibfnamefont {R.~J.}\ \bibnamefont {Cava}},\ }\href {\doibase
  10.1103/PhysRevB.81.195203} {\bibfield  {journal} {\bibinfo  {journal} {Phys.
  Rev. B}\ }\textbf {\bibinfo {volume} {81}},\ \bibinfo {pages} {195203}
  (\bibinfo {year} {2010})}\BibitemShut {NoStop}%
\bibitem [{\citenamefont {Xu}\ \emph {et~al.}(2012)\citenamefont {Xu},
  \citenamefont {Neupane}, \citenamefont {Liu}, \citenamefont {Zhang},
  \citenamefont {Richardella}, \citenamefont {Wray}, \citenamefont {Alidoust},
  \citenamefont {Leandersson}, \citenamefont {Balasubramanian}, \citenamefont
  {S{\'a}nchez-Barriga} \emph {et~al.}}]{Xu2012}%
  \BibitemOpen
  \bibfield  {author} {\bibinfo {author} {\bibfnamefont {S.-Y.}\ \bibnamefont
  {Xu}}, \bibinfo {author} {\bibfnamefont {M.}~\bibnamefont {Neupane}},
  \bibinfo {author} {\bibfnamefont {C.}~\bibnamefont {Liu}}, \bibinfo {author}
  {\bibfnamefont {D.}~\bibnamefont {Zhang}}, \bibinfo {author} {\bibfnamefont
  {A.}~\bibnamefont {Richardella}}, \bibinfo {author} {\bibfnamefont {L.~A.}\
  \bibnamefont {Wray}}, \bibinfo {author} {\bibfnamefont {N.}~\bibnamefont
  {Alidoust}}, \bibinfo {author} {\bibfnamefont {M.}~\bibnamefont
  {Leandersson}}, \bibinfo {author} {\bibfnamefont {T.}~\bibnamefont
  {Balasubramanian}}, \bibinfo {author} {\bibfnamefont {J.}~\bibnamefont
  {S{\'a}nchez-Barriga}},  \emph {et~al.},\ }\href@noop {} {\bibfield
  {journal} {\bibinfo  {journal} {Nature Physics}\ }\textbf {\bibinfo {volume}
  {8}},\ \bibinfo {pages} {616} (\bibinfo {year} {2012})}\BibitemShut {NoStop}%
\bibitem [{\citenamefont {Shelford}\ \emph {et~al.}(2012)\citenamefont
  {Shelford}, \citenamefont {Hesjedal}, \citenamefont {Collins-McIntyre},
  \citenamefont {Dhesi}, \citenamefont {Maccherozzi},\ and\ \citenamefont
  {van~der Laan}}]{PhysRevB.86.081304}%
  \BibitemOpen
  \bibfield  {author} {\bibinfo {author} {\bibfnamefont {L.~R.}\ \bibnamefont
  {Shelford}}, \bibinfo {author} {\bibfnamefont {T.}~\bibnamefont {Hesjedal}},
  \bibinfo {author} {\bibfnamefont {L.}~\bibnamefont {Collins-McIntyre}},
  \bibinfo {author} {\bibfnamefont {S.~S.}\ \bibnamefont {Dhesi}}, \bibinfo
  {author} {\bibfnamefont {F.}~\bibnamefont {Maccherozzi}}, \ and\ \bibinfo
  {author} {\bibfnamefont {G.}~\bibnamefont {van~der Laan}},\ }\href {\doibase
  10.1103/PhysRevB.86.081304} {\bibfield  {journal} {\bibinfo  {journal} {Phys.
  Rev. B}\ }\textbf {\bibinfo {volume} {86}},\ \bibinfo {pages} {081304}
  (\bibinfo {year} {2012})}\BibitemShut {NoStop}%
\bibitem [{\citenamefont {Li}\ \emph {et~al.}(2012)\citenamefont {Li},
  \citenamefont {Yang}, \citenamefont {Chen}, \citenamefont {Whangbo},
  \citenamefont {Xiang},\ and\ \citenamefont {Gong}}]{li2012strong}%
  \BibitemOpen
  \bibfield  {author} {\bibinfo {author} {\bibfnamefont {Z.~L.}\ \bibnamefont
  {Li}}, \bibinfo {author} {\bibfnamefont {J.~H.}\ \bibnamefont {Yang}},
  \bibinfo {author} {\bibfnamefont {G.~H.}\ \bibnamefont {Chen}}, \bibinfo
  {author} {\bibfnamefont {M.-H.}\ \bibnamefont {Whangbo}}, \bibinfo {author}
  {\bibfnamefont {H.~J.}\ \bibnamefont {Xiang}}, \ and\ \bibinfo {author}
  {\bibfnamefont {X.~G.}\ \bibnamefont {Gong}},\ }\href {\doibase
  10.1103/PhysRevB.85.054426} {\bibfield  {journal} {\bibinfo  {journal} {Phys.
  Rev. B}\ }\textbf {\bibinfo {volume} {85}},\ \bibinfo {pages} {054426}
  (\bibinfo {year} {2012})}\BibitemShut {NoStop}%
\bibitem [{\citenamefont {Zhang}\ \emph {et~al.}(2012)\citenamefont {Zhang},
  \citenamefont {Zhu}, \citenamefont {Zhang}, \citenamefont {Xiao},\ and\
  \citenamefont {Yao}}]{zhang2012tailoring}%
  \BibitemOpen
  \bibfield  {author} {\bibinfo {author} {\bibfnamefont {J.-M.}\ \bibnamefont
  {Zhang}}, \bibinfo {author} {\bibfnamefont {W.}~\bibnamefont {Zhu}}, \bibinfo
  {author} {\bibfnamefont {Y.}~\bibnamefont {Zhang}}, \bibinfo {author}
  {\bibfnamefont {D.}~\bibnamefont {Xiao}}, \ and\ \bibinfo {author}
  {\bibfnamefont {Y.}~\bibnamefont {Yao}},\ }\href@noop {} {\bibfield
  {journal} {\bibinfo  {journal} {Phys. Rev. Lett.}\ }\textbf {\bibinfo
  {volume} {109}},\ \bibinfo {pages} {266405} (\bibinfo {year}
  {2012})}\BibitemShut {NoStop}%
\bibitem [{\citenamefont {Henk}\ \emph
  {et~al.}(2012{\natexlab{b}})\citenamefont {Henk}, \citenamefont {Ernst},
  \citenamefont {Eremeev}, \citenamefont {Chulkov}, \citenamefont
  {Maznichenko},\ and\ \citenamefont {Mertig}}]{henk2012complex}%
  \BibitemOpen
  \bibfield  {author} {\bibinfo {author} {\bibfnamefont {J.}~\bibnamefont
  {Henk}}, \bibinfo {author} {\bibfnamefont {A.}~\bibnamefont {Ernst}},
  \bibinfo {author} {\bibfnamefont {S.~V.}\ \bibnamefont {Eremeev}}, \bibinfo
  {author} {\bibfnamefont {E.~V.}\ \bibnamefont {Chulkov}}, \bibinfo {author}
  {\bibfnamefont {I.~V.}\ \bibnamefont {Maznichenko}}, \ and\ \bibinfo {author}
  {\bibfnamefont {I.}~\bibnamefont {Mertig}},\ }\href {\doibase
  10.1103/PhysRevLett.108.206801} {\bibfield  {journal} {\bibinfo  {journal}
  {Phys. Rev. Lett.}\ }\textbf {\bibinfo {volume} {108}},\ \bibinfo {pages}
  {206801} (\bibinfo {year} {2012}{\natexlab{b}})}\BibitemShut {NoStop}%
\bibitem [{\citenamefont {Schmidt}\ \emph {et~al.}(2011)\citenamefont
  {Schmidt}, \citenamefont {Miwa},\ and\ \citenamefont
  {Fazzio}}]{schmidt2011spin}%
  \BibitemOpen
  \bibfield  {author} {\bibinfo {author} {\bibfnamefont {T.~M.}\ \bibnamefont
  {Schmidt}}, \bibinfo {author} {\bibfnamefont {R.~H.}\ \bibnamefont {Miwa}}, \
  and\ \bibinfo {author} {\bibfnamefont {A.}~\bibnamefont {Fazzio}},\ }\href
  {\doibase 10.1103/PhysRevB.84.245418} {\bibfield  {journal} {\bibinfo
  {journal} {Phys. Rev. B}\ }\textbf {\bibinfo {volume} {84}},\ \bibinfo
  {pages} {245418} (\bibinfo {year} {2011})}\BibitemShut {NoStop}%
\bibitem [{\citenamefont {Biswas}\ and\ \citenamefont
  {Balatsky}(2010)}]{PhysRevB.81.233405}%
  \BibitemOpen
  \bibfield  {author} {\bibinfo {author} {\bibfnamefont {R.~R.}\ \bibnamefont
  {Biswas}}\ and\ \bibinfo {author} {\bibfnamefont {A.~V.}\ \bibnamefont
  {Balatsky}},\ }\href {\doibase 10.1103/PhysRevB.81.233405} {\bibfield
  {journal} {\bibinfo  {journal} {Phys. Rev. B}\ }\textbf {\bibinfo {volume}
  {81}},\ \bibinfo {pages} {233405} (\bibinfo {year} {2010})}\BibitemShut
  {NoStop}%
\bibitem [{\citenamefont {Abanin}\ and\ \citenamefont
  {Pesin}(2011)}]{PhysRevLett.106.136802}%
  \BibitemOpen
  \bibfield  {author} {\bibinfo {author} {\bibfnamefont {D.~A.}\ \bibnamefont
  {Abanin}}\ and\ \bibinfo {author} {\bibfnamefont {D.~A.}\ \bibnamefont
  {Pesin}},\ }\href {\doibase 10.1103/PhysRevLett.106.136802} {\bibfield
  {journal} {\bibinfo  {journal} {Phys. Rev. Lett.}\ }\textbf {\bibinfo
  {volume} {106}},\ \bibinfo {pages} {136802} (\bibinfo {year}
  {2011})}\BibitemShut {NoStop}%
\bibitem [{\citenamefont {Efimkin}\ and\ \citenamefont
  {Galitski}(2014)}]{PhysRevB.89.115431}%
  \BibitemOpen
  \bibfield  {author} {\bibinfo {author} {\bibfnamefont {D.~K.}\ \bibnamefont
  {Efimkin}}\ and\ \bibinfo {author} {\bibfnamefont {V.}~\bibnamefont
  {Galitski}},\ }\href {\doibase 10.1103/PhysRevB.89.115431} {\bibfield
  {journal} {\bibinfo  {journal} {Phys. Rev. B}\ }\textbf {\bibinfo {volume}
  {89}},\ \bibinfo {pages} {115431} (\bibinfo {year} {2014})}\BibitemShut
  {NoStop}%
\bibitem [{\citenamefont {Choi}\ \emph {et~al.}(2013)\citenamefont {Choi},
  \citenamefont {Jo}, \citenamefont {Lee}, \citenamefont {Lee}, \citenamefont
  {Jo}, \citenamefont {Kajino}, \citenamefont {Takabatake}, \citenamefont {Ko},
  \citenamefont {Park},\ and\ \citenamefont {Jung}}]{APL_152103}%
  \BibitemOpen
  \bibfield  {author} {\bibinfo {author} {\bibfnamefont {Y.~H.}\ \bibnamefont
  {Choi}}, \bibinfo {author} {\bibfnamefont {N.~H.}\ \bibnamefont {Jo}},
  \bibinfo {author} {\bibfnamefont {K.~J.}\ \bibnamefont {Lee}}, \bibinfo
  {author} {\bibfnamefont {H.~W.}\ \bibnamefont {Lee}}, \bibinfo {author}
  {\bibfnamefont {Y.~H.}\ \bibnamefont {Jo}}, \bibinfo {author} {\bibfnamefont
  {J.}~\bibnamefont {Kajino}}, \bibinfo {author} {\bibfnamefont
  {T.}~\bibnamefont {Takabatake}}, \bibinfo {author} {\bibfnamefont {K.-T.}\
  \bibnamefont {Ko}}, \bibinfo {author} {\bibfnamefont {J.-H.}\ \bibnamefont
  {Park}}, \ and\ \bibinfo {author} {\bibfnamefont {M.~H.}\ \bibnamefont
  {Jung}},\ }\href@noop {} {\bibfield  {journal} {\bibinfo  {journal} {Appl.
  Phys. Lett.}\ }\textbf {\bibinfo {volume} {25}},\ \bibinfo {pages} {206005}
  (\bibinfo {year} {2013})}\BibitemShut {NoStop}%
\bibitem [{\citenamefont {Jan{\'\i}{\v{c}}ek}\ \emph
  {et~al.}(2008)\citenamefont {Jan{\'\i}{\v{c}}ek}, \citenamefont
  {Dra{\v{s}}ar}, \citenamefont {Lo{\v{s}}t{\'a}k}, \citenamefont
  {Vejpravov{\'a}},\ and\ \citenamefont {Sechovsk{\`y}}}]{phys_20083553}%
  \BibitemOpen
  \bibfield  {author} {\bibinfo {author} {\bibfnamefont {P.}~\bibnamefont
  {Jan{\'\i}{\v{c}}ek}}, \bibinfo {author} {\bibfnamefont
  {{\v{C}}.}~\bibnamefont {Dra{\v{s}}ar}}, \bibinfo {author} {\bibfnamefont
  {P.}~\bibnamefont {Lo{\v{s}}t{\'a}k}}, \bibinfo {author} {\bibfnamefont
  {J.}~\bibnamefont {Vejpravov{\'a}}}, \ and\ \bibinfo {author} {\bibfnamefont
  {V.}~\bibnamefont {Sechovsk{\`y}}},\ }\href {\doibase
  http://dx.doi.org/10.1016/j.physb.2008.05.025} {\bibfield  {journal}
  {\bibinfo  {journal} {Physica B: Condensed Matter}\ }\textbf {\bibinfo
  {volume} {403}},\ \bibinfo {pages} {3553 } (\bibinfo {year}
  {2008})}\BibitemShut {NoStop}%
\bibitem [{\citenamefont {Mahani}\ \emph {et~al.}(2014)\citenamefont {Mahani},
  \citenamefont {Islam}, \citenamefont {Pertsova},\ and\ \citenamefont
  {Canali}}]{rm_d_level}%
  \BibitemOpen
  \bibfield  {author} {\bibinfo {author} {\bibfnamefont {M.~R.}\ \bibnamefont
  {Mahani}}, \bibinfo {author} {\bibfnamefont {M.~F.}\ \bibnamefont {Islam}},
  \bibinfo {author} {\bibfnamefont {A.}~\bibnamefont {Pertsova}}, \ and\
  \bibinfo {author} {\bibfnamefont {C.~M.}\ \bibnamefont {Canali}},\ }\href
  {\doibase 10.1103/PhysRevB.89.165408} {\bibfield  {journal} {\bibinfo
  {journal} {Phys. Rev. B}\ }\textbf {\bibinfo {volume} {89}},\ \bibinfo
  {pages} {165408} (\bibinfo {year} {2014})}\BibitemShut {NoStop}%
\bibitem [{\citenamefont {Anisimov}\ \emph {et~al.}(1991)\citenamefont
  {Anisimov}, \citenamefont {Zaanen},\ and\ \citenamefont
  {Andersen}}]{PhysRevB.44.943}%
  \BibitemOpen
  \bibfield  {author} {\bibinfo {author} {\bibfnamefont {V.~I.}\ \bibnamefont
  {Anisimov}}, \bibinfo {author} {\bibfnamefont {J.}~\bibnamefont {Zaanen}}, \
  and\ \bibinfo {author} {\bibfnamefont {O.~K.}\ \bibnamefont {Andersen}},\
  }\href {\doibase 10.1103/PhysRevB.44.943} {\bibfield  {journal} {\bibinfo
  {journal} {Phys. Rev. B}\ }\textbf {\bibinfo {volume} {44}},\ \bibinfo
  {pages} {943} (\bibinfo {year} {1991})}\BibitemShut {NoStop}%
\bibitem [{\citenamefont {Blaha}\ \emph {et~al.}(2001)\citenamefont {Blaha},
  \citenamefont {Schwarz}, \citenamefont {Madsen}, \citenamefont {Kvasnicka},\
  and\ \citenamefont {Luitz}}]{Wien2k_package}%
  \BibitemOpen
  \bibfield  {author} {\bibinfo {author} {\bibfnamefont {P.}~\bibnamefont
  {Blaha}}, \bibinfo {author} {\bibfnamefont {K.}~\bibnamefont {Schwarz}},
  \bibinfo {author} {\bibfnamefont {G.~K.~H.}\ \bibnamefont {Madsen}}, \bibinfo
  {author} {\bibfnamefont {D.}~\bibnamefont {Kvasnicka}}, \ and\ \bibinfo
  {author} {\bibfnamefont {J.}~\bibnamefont {Luitz}},\ }\href@noop {} {\emph
  {\bibinfo {title} {WIEN2k, An Augmented Plane Wave Plus Local Orbitals
  Program for Calculating Crystal properties (Vienna University of Technology,
  Austria)}}} (\bibinfo {year} {2001})\BibitemShut {NoStop}%
\bibitem [{\citenamefont {Perdew}\ \emph {et~al.}(1996)\citenamefont {Perdew},
  \citenamefont {Burke},\ and\ \citenamefont {Ernzerhof}}]{perdew96}%
  \BibitemOpen
  \bibfield  {author} {\bibinfo {author} {\bibfnamefont {J.~P.}\ \bibnamefont
  {Perdew}}, \bibinfo {author} {\bibfnamefont {K.}~\bibnamefont {Burke}}, \
  and\ \bibinfo {author} {\bibfnamefont {M.}~\bibnamefont {Ernzerhof}},\ }\href
  {\doibase 10.1103/PhysRevLett.77.3865} {\bibfield  {journal} {\bibinfo
  {journal} {Phys. Rev. Lett.}\ }\textbf {\bibinfo {volume} {77}},\ \bibinfo
  {pages} {3865} (\bibinfo {year} {1996})}\BibitemShut {NoStop}%
\bibitem [{\citenamefont {Kobayashi}(2011)}]{kobayashi2011electron}%
  \BibitemOpen
  \bibfield  {author} {\bibinfo {author} {\bibfnamefont {K.}~\bibnamefont
  {Kobayashi}},\ }\href {\doibase 10.1103/PhysRevB.84.205424} {\bibfield
  {journal} {\bibinfo  {journal} {Phys. Rev. B}\ }\textbf {\bibinfo {volume}
  {84}},\ \bibinfo {pages} {205424} (\bibinfo {year} {2011})}\BibitemShut
  {NoStop}%
\bibitem [{\citenamefont {Pertsova}\ and\ \citenamefont
  {Canali}(2014)}]{ap_TI_prob}%
  \BibitemOpen
  \bibfield  {author} {\bibinfo {author} {\bibfnamefont {A.}~\bibnamefont
  {Pertsova}}\ and\ \bibinfo {author} {\bibfnamefont {C.}~\bibnamefont
  {Canali}},\ }\href@noop {} {\bibfield  {journal} {\bibinfo  {journal} {New J.
  Phys}\ }\textbf {\bibinfo {volume} {16}},\ \bibinfo {pages} {063022}
  (\bibinfo {year} {2014})}\BibitemShut {NoStop}%
\bibitem [{\citenamefont {Chapler}\ \emph {et~al.}(2014)\citenamefont
  {Chapler}, \citenamefont {Post}, \citenamefont {Richardella}, \citenamefont
  {Lee}, \citenamefont {Tao}, \citenamefont {Samarth},\ and\ \citenamefont
  {Basov}}]{PhysRevB.89.235308}%
  \BibitemOpen
  \bibfield  {author} {\bibinfo {author} {\bibfnamefont {B.~C.}\ \bibnamefont
  {Chapler}}, \bibinfo {author} {\bibfnamefont {K.~W.}\ \bibnamefont {Post}},
  \bibinfo {author} {\bibfnamefont {A.~R.}\ \bibnamefont {Richardella}},
  \bibinfo {author} {\bibfnamefont {J.~S.}\ \bibnamefont {Lee}}, \bibinfo
  {author} {\bibfnamefont {J.}~\bibnamefont {Tao}}, \bibinfo {author}
  {\bibfnamefont {N.}~\bibnamefont {Samarth}}, \ and\ \bibinfo {author}
  {\bibfnamefont {D.~N.}\ \bibnamefont {Basov}},\ }\href {\doibase
  10.1103/PhysRevB.89.235308} {\bibfield  {journal} {\bibinfo  {journal} {Phys.
  Rev. B}\ }\textbf {\bibinfo {volume} {89}},\ \bibinfo {pages} {235308}
  (\bibinfo {year} {2014})}\BibitemShut {NoStop}%
\end{thebibliography}%
\end{document}